\documentclass[11pt,oneside, letter]{article}

\usepackage{amsthm, amsmath, amssymb, amsfonts, graphicx, epsfig}
\usepackage{accents}

\usepackage{bbm}
\usepackage{mathrsfs}

\usepackage{cancel}

\usepackage{epstopdf}

\usepackage{graphicx}
\usepackage[font=sl,labelfont=bf]{caption}
\usepackage{subcaption}

\usepackage{float}
\restylefloat{table}

\usepackage{enumerate}

\usepackage[usenames,dvipsnames]{color}

\usepackage[ulem=normalem,draft]{changes}       

\usepackage[breaklinks=true,colorlinks=true, pdfstartview=FitV, linkcolor=blue,
            citecolor=blue, urlcolor=blue]{hyperref}
\usepackage[usenames]{color}
\usepackage{breakcites}

\usepackage{url}
\makeatletter\def\url@leostyle{%
 \@ifundefined{selectfont}{\def\UrlFont{\sf}}{\def\UrlFont{\scriptsize\ttfamily}}} \makeatother

\usepackage[framemethod=default]{mdframed}

\usepackage{array}

\usepackage{booktabs}

\usepackage[margin=1.0in, letterpaper]{geometry}

\newtheorem{theorem}{Theorem}
\newtheorem{proposition}[theorem]{Proposition}

\theoremstyle{definition}

\theoremstyle{remark}
\newtheorem{remark}[theorem]{Remark}

\numberwithin{equation}{section}
\numberwithin{theorem}{section}

\definecolor{Red}{rgb}{0.9,0,0.0}
\definecolor{Blue}{rgb}{0,0.0,1.0}

\def\bN{\mathbb{N}}

\def\bR{\mathbb{R}}

\newcommand{\1}{\mathbf{1}}            

\DeclareMathOperator{\var}{\mathrm{V}@\mathrm{R}}           

\title{Backtesting Expected Shortfall: a simple recipe?\,\thanks{The authors express their gratitude to the anonymous referee for valuable comments and suggestions that helped to improve the paper. We thank S\'ebastien Ray for stimulating discussions and very helpful remarks. Part of the work of the second author was supported by the National Science Centre, Poland, via project 2016/23/B/ST1/00479.\vspace{0.5em}}
$^{,}$\thanks{The views and opinions expressed in this article are the authors' own and do not necessarily reflect the views and opinions of their current or past employers.
\vspace{0.5em}}}

\makeatletter
\def\and{%
  \end{tabular}%
  \begin{tabular}[t]{c}}%
\def\@fnsymbol#1{\ensuremath{\ifcase#1\or a\or b\or c\or
   d\or e\or f\or g\or h\or i\else\@ctrerr\fi}}
\makeatother

\author{
         Felix Moldenhauer\,\thanks{HSBC Global Banking and Markets, London, United Kingdom
         \newline \hspace*{1.45em} E-mail: \url{felix.moldenhauer@hsbcib.com}
         \vspace{0.5em}} ,
\and
	Marcin Pitera\,\thanks{ Institute of Mathematics, Jagiellonian University, Cracow, Poland
         \newline \hspace*{1.45em}  E-mail: \url{marcin.pitera@uj.edu.pl},  URL: \url{http://www2.im.uj.edu.pl/MarcinPitera/}
          }
        }

\date{ {\small This version: \today}} 

\usepackage{natbib}
\defcitealias{Bas2013}{BCBS, 2013} 
\defcitealias{Bas2011}{BCBS, 2011}
\defcitealias{Bas2006}{BCBS, 2006} 
\defcitealias{Bas1996}{BCBS, 1996}   
\defcitealias{Bas2016}{BCBS, 2016}   

\newcommand{\Ind}{{\mathbf 1}}
\newcommand{\ind}[1]{\Ind_{\{#1\}}}

\usepackage{doi}
\usepackage{placeins}

\begin{document}

\maketitle

{\footnotesize
\begin{tabular}{l@{} p{350pt}}
  \hline \\[-.2em]
  \textsc{Abstract}: \ & We propose a new backtesting framework for Expected Shortfall that could be used by the regulator. Instead of looking at the estimated capital reserve and the realised cash-flow separately, one could bind them into the {\it secured position}, for which risk measurement is much easier. Using this simple concept combined with monotonicity of Expected Shortfall with respect to its target confidence level we introduce a natural and efficient backtesting framework. Our test statistics is given by the biggest number of worst realisations for the secured position that add up to a negative total. Surprisingly, this simple quantity could be used to construct an efficient backtesting framework for unconditional coverage of Expected Shortfall in a natural extension of the regulatory traffic-light approach for Value-at-Risk. While being easy to calculate, the test statistic is based on the underlying duality between coherent risk measures and scale-invariant performance measures.
 \\[0.5em]
\textsc{Keywords:} \ & 
 value-at-risk, expected shortfall, backtesting, backtest, risk bias, risk estimation, risk conservativeness, internal model-based approach, unconditional coverage test, fundamental review of the trading book\\
\textsc{JEL:} \ & C19, C50, D81, G17, G28  \\[1em]
  \hline
\end{tabular}
}

\section{Introduction}\label{S:intro}

Risk measures play a major role in the computation of regulatory capital that is required to ensure financial stability of the underlying financial institution. Because of that, the regulator needs to ensure that the risk estimation methodology adopted by the institution is conservative and that the resulting capital reserves are robust; see~\cite{Car2009},~\cite{MFE}, and references therein.

The backtesting procedure is one of the key quantitative tools used by the regulators to assess the conservativeness of the risk measurement methodology. Because of that, the backtests as well as the related statistical properties of risk estimators are intensively studied and the estimation techniques are being constantly improved; see \cite{Davis2016,ConDegSca2010,Acerbi2014Risk,Ziegel2014,Frank2016} for exemplary recent contributions.

Currently, there is an intensive debate about two risk measures: Value-at-Risk (VAR) and Expected Shortfall (ES). The discussion is stimulated by recent regulatory developments (e.g. FRTB, ICS) as well as the propagation of elicitability concept in academia; see~\cite{AceSze2017} and references therein. In particular, the update of Basel's capital requirements for market risk under the internal model-based approach (IMA) raised a lot of concern because of the replacement of VAR at level $1\%$ with ES at level $2.5\%$; see \citepalias{Bas2016}. Using regulatory traffic-light backtest based on the exception rate for ES is highly criticised as such framework is inconsistent with the underlying risk measurement philosophy: counting quantile breaches is strictly related to the VAR risk measure.

Also, the more fundamental question of whether ES is even possible to backtest has been asked. In particular, \cite{Gneiting2011} showed that ES is not elicitable; see also \cite{Web2006}. Following this finding, many others have interpreted this as the evidence that it is not possible to efficiently backtest ES at all; see e.g.~\cite{Car2013}. On the other hand, in~\cite{Acerbi2014Risk} it is stated that elicitability is connected to model comparison rather than to model testing, so that the lack of elicitability is not crucial when backtesting is considered. Moreover, it was shown in~\cite{FisZie2016} that ES is jointly elicitable with VAR so that the elicitability testing techniques could be adopted for ES; see~\cite{NolZie2017}. For further details see \cite{Car2014,Emmeretal2015,Davis2016,AceSze2017} and references therein.

Finally, it should be noted that the definition/concept of {\it backtesting} is not uniform so that the statements like ''backtesting of ES is (not) possible'' are merely subjective expressions rather than scientific facts. 

In the literature, a lot of possible backtesting frameworks have been proposed, most of them not directly linked to elicitability. In particular,~\cite{Won2008} propose a parametric saddle-point method,~\cite{RigCar2013} studies the truncation based ES backtests,~\cite{Emmeretal2015} approximate ES using VARs with different levels and use the standard backtesting tools, and  \cite{Acerbi2014Risk} propose a backtest based on a specific ES normalisation. We refer to~\cite{AceSze2017,NolZie2017,FisZieGne2017} for various backtesting procedures based on elicitability, and to~\cite{McnAleFre2000,Ber2001,KerMel2004,CosCur2015,DuEsc2016,LosWieZig2018} for other alternative methods.

Unfortunately, most of the methods mentioned in the previous paragraphs require advanced mathematical framework, certain model assumptions, reference estimation process, and/or relatively large samples. The lack of transparent and straightforward backtesting framework that could be used for any IMA model, and for which the output would be self-explanatory in the financial context is rather surprising, especially given the simplicity and elegance of the exception-rate procedure used for VAR backtesting; see~\citepalias{Bas1996}.


In this paper, we try to remedy this problem by proposing a new ES backtesting framework. We focus on unconditional coverage backtesting, as the independence of the reserve-capital breaks is typically assessed by visual inspections. Instead of looking at the estimated capital reserve and the realised financial position cash-flow separately, we bind them into the {\it secured position}, for which risk measurement is much easier. This simple and intuitive observation leads to risk bias concept that was recently developed in~\cite{PitSch2016}.

Using this framework, we propose a natural backtest for ES by focussing on how many of the secured position's worst realisations (during a given period) still sum to a negative total. In contrast, VaR backtests count the number of negative realisations of the secured position, called {\it breaches}.

In analogy to the current regulatory framework, traffic-light thresholds with associated capital penalties can be set against our 'largest number of worst realisations yielding a negative sum', preserving its positive (and negative) properties. We believe that the proposed framework is simple, self-explanatory, and efficient.

Moreover, we show that our test statistic is a performance measure obtained using the duality relationship between coherent risk measures and scale-invariant performance measures; see Equality~\eqref{eq:es.hist}. We also show that in the VAR framework the similar relationship is true when the exception rate statistic is considered. This explains the tight connections between VAR and exception rate test, and provides mathematical justification for the choice of our test statistic.

It should be noted that given a sample with realisations of the secured position our test statistic could be easily implemented in almost any programming language. For example, in {\bf R} software, assuming we get the secured position sample vector $y$, the biggest number of worst realisations that add up to a negative total could be computed in one line of code: {sum(cumsum(sort(y))$<$0)}.

This note is organised as follows: In Section~\ref{S:preliminaries}, we provide the background to our framework. Section~\ref{S:var.bt} is dedicated to the description of the standard regulatory VAR backtest. In particular, we show the connections to our framework, discuss the concept of bias in VAR model, and explain why the test statistic could be interpreted as an acceptability index. Section~\ref{S:es.bt} contains the main contribution of this paper, i.e. the description of the ES backtesting framework, while in Section~\ref{S:test.performance} we investigate the distribution of the proposed test statistic. Next, in Section~\ref{Sec:empirical} we provide a small empirical study to show how the proposed framework perform on market and simulated data. In particular, we analyse how the empirical results for ES backtest are aligned with the results of the classical regulatory VAR backtest. We conclude in Section~\ref{Sec:conclude}.

\section{Preliminaries}\label{S:preliminaries}
In this section we outline the idea of risk unbiasedness and introduce related notation; for a rigorous mathematical framework and formal definitions we refer to~\cite{PitSch2016}.

Let $\rho$ be a distribution-based risk measure (e.g. VAR or ES). We assume that we have an Internal Model  (IM) that is used to compute the capital reserve to protect against fluctuations in the future value of a financial portfolio. For example, this could refer to Historical Simulation, Gaussian, or Monte Carlo risk estimation models; see~\cite{Car2009}. For transparency, we assume that the holding period is 1-day and use $\textrm{P\&L}$ to denote a random variable associated with the future portfolio profits and losses. We use $\hat\rho$ to denote the estimator of $\rho(\textrm{P\&L})$ that was computed using historical data combined with the IM methodology, i.e. $\hat\rho$ determines the estimated capital reserve for P\&L.

Instead of independently calculating the (theoretical) risk of $\textrm{P\&L}$ and then comparing it to the IM estimator $\hat\rho$ we can bind those values into a single random variable
\[
Y:=\textrm{P\&L}+\hat\rho,
\]
that we call the {\it secured position}. In particular, if $Y$ is positive, it implies that the estimated capital reserve is high enough to cover the (occurred) loss. Given risk measure $\rho$, we are interested in measuring the riskiness of the secured position $Y$. If the risk of $Y$ is equal to zero, that is if
\begin{equation}\label{eq:zero.risk}
\rho(Y)=0,
\end{equation}
then the estimated capital reserve might be considered sufficient. In fact, if we knew the theoretical risk of $\textrm{P\&L}$, then we could simply set $\hat\rho=\rho(\textrm{P\&L})$, and using the cash-additivity property get
\[
\rho(Y)=\rho(\textrm{P\&L}+\rho(\textrm{P\&L}))=\rho(\textrm{P\&L})-\rho(\textrm{P\&L})=0.
\]
Unfortunately, the quantity $\rho(\textrm{P\&L})$ is usually not known and needs to be estimated, for example by using historical data. Consequently, $\hat\rho$ becomes a random variable that depends on estimation procedure, and Equality~\eqref{eq:zero.risk} is much harder to obtain; see~\cite{PitSch2016} for details.

While from the financial institution's perspective it might be optimal to find $\hat\rho$ for which the risk of $Y$ is zero, the regulator is more interested whether the position $Y$ is {\it acceptable}, i.e.
\begin{equation}\label{eq:cons}
\rho(Y)\leq 0.
\end{equation}
In other words, rather than checking the overall performance of IM risk estimator $\hat\rho$, we are interested in the conservativeness of the model: if $\rho(Y)<0$, then the estimator $\hat\rho$ overestimates the risk of $\textrm{P\&L}$. In that case, the regulator is unlikely to reject the methodology as the resulting capital reserve is higher than required by the portfolio/position's risk.\footnote{The situation is not symmetrical -- we penalise only the underestimation of risk. The backtests that are based on risk estimator overall fit rather than conservativeness are not in line with our definition. Note that regulations are consistent with the concept of conservativeness: VAR models with no breaches are in the green zone.}

Let us now introduce the basic notation that will be used in the following sections. We use $n\in\bN$ to denote the number of days used in the backtest, i.e. the length of the backtesting window. For $i=1,\ldots,n$\, we denote by $\textrm{P\&L}_i$ portfolio's P\&L realised on day $i$ and we use $\hat\rho_i$ to denote the corresponding capital reserve that was estimated using the IM methodology (for VAR or ES) combined with data available up to day $i-1$. Next, we use
\begin{equation}\label{eq:secures.position}
y_i:=\textrm{P\&L}_{i}+\hat\rho_i
\end{equation}
to denote the $i$-th day realisation of the secured position and $y:=(y_1,\ldots,y_n)$ to denote the vector of realisations during the backtest window.

Being the sum of P\&L realisation and estimated capital reserve, the secured position inherits the statistical properties of both. It is well known that returns in financial markets are hard to predict due to changing volatility patterns; see \cite{Man1963}. While the capital reserves are supposed to change in line with the riskiness of the protected portfolio, in practice they tend to adjust rather slowly, especially if estimated on the basis of historical data series.

What does the combination of uncorrelated, but heteroscedastic $(\textrm{P\&L}_i)$ with highly autocorrelated $(\hat \rho_i)$ imply for $y$?  We argue that statistical robustness of the resulting secured position is increased for the left/loss tail, since adjustments in the capital reserve partially offset changing P\&L volatility. As the combined position remains largely uncorrelated due to randomness in P\&L dominating the rather slow adjustments in the estimated risk capital reserve, the vector $y$ gets closer to meeting the i.i.d. assumption -- in the sense required by our backtests. We provide detailed discussions of this claim in reference to VAR and ES backtesting in Section~\ref{S:var.bt} and Section~\ref{S:es.bt}, respectively.

In the following sections, we assume i.i.d. behaviour for the relevant part of the secured position vector $y$, implicitly relying on a correct model specification for estimating the capital reserve. To explicitly account for changes in the risk profile of the portfolio and/or volatility clustering, one might consider appropriate normalisation schemes for the P\&L vector $y$; see Remark~\ref{rem:norm} for details.

\begin{remark}\label{rem:norm} 
If the portfolio risk-profile is changing, then the realised values of the secured portfolio position given in \eqref{eq:secures.position} might not accurately reflect the current risk exposure. To mitigate that effect one might normalise the risk values and consider the modified secured position sample
\begin{equation}\label{eq:normalisation}
\widetilde y_i:=\tfrac{\textrm{P\&L}_{i}+\hat\rho_i}{\hat\rho_i}=\tfrac{\textrm{P\&L}_{i}}{\hat\rho_i}+1.
\end{equation}
In simple words, we re-scale the realised P\&Ls so that the estimated risk is constant and equal to 1. While \eqref{eq:secures.position} calls for constant volatility, risk in \eqref{eq:normalisation} is proportional to portfolio volatility -- given correct model specification for $\hat \rho_i$. This allows a time-consistent risk measurement of the secured position; such operation is rational if the underlying risk measure is positively homogenous allowing linear transformation of risk. Note that this transform has been used in ES backtesting context in~\cite{Acerbi2014Risk}. Alternatively, we might normalise $y$ by dividing by the total portfolio value, i.e. consider return rates instead of P\&Ls. However, this might be problematic as in practice we often have zero-worth portfolio with material risk, and such transform does not take into account the varying volatility effect. The proposed transforms are especially important for real-life trading desks when the actual P\&Ls are confronted against historical risk estimators; for hypothetical portfolio backtests the standard  secured position~\eqref{eq:secures.position} could be used.
\end{remark}

\section{Value-at-Risk backtesting}\label{S:var.bt}

The typical VaR backtesting framework tests unconditional coverage and is based on the exception rate (breach) test; see e.g. \cite{Car2009}. For $y$ corresponding to a sample of secured position backed by capital amounting to VAR, the exception rate backtest statistic is given by
\begin{equation}\label{eq:Tn.first}
T_n:=\sum_{i=1}^{n} \frac{\1_{\{y_i<0\}}}{n},
\end{equation}
where $\1_A$ is the indicator function of set $A$. Simply speaking, we count the number of exceptions (capital breaks) in the secured sample and divide it by $n$ to get the average exception rate.

In \eqref{eq:Tn.first}, we link the correct specification of the underlying VAR model to its property of producing the correct Bernoulli distributed number of breaches. This is significantly less restrictive than imposing an i.i.d. assumption on the P\&L vector used in the historical VAR estimator; see Section~\ref{S:preliminaries} for the discussion. In particular, note that in the VAR case the normalisation scheme proposed in Remark~\ref{rem:norm} does not have an impact on the value of $T_n$ as the breach is determined only by the sign of $y_i$; it is easy to show that $\1_{\{\widetilde y_i<0\}}=\1_{\{y_i<0\}}$ which in turn implies invariance of $T_n$ with respect to the normalisation scheme.

To assess the performance of an internal model for market risk capital calculations, the regulator defines three zones according to the number of backtesting breaches observed during a year. A correctly specified model for the 1\% VAR reference risk metric is expected to produce 2 or 3 breaches during $n=250$ (business) days. Constructing a confidence band around the number of expected breaches, an internal model (IM) is said to be in:
\begin{enumerate}[-]
\item {\bf green zone}, if there are less then 5 breaches: under the correct model, this is expected to happen in around 90\% of all cases and corresponds to $T_n \in[0.00, 0.02)$;

\item {\bf yellow zone}, if the number of breaches is between 5 and 9: under the correct model, this is expected to happen in around 10\% of all cases and corresponds to $T_n\in [0.02, 0.04)$;

\item {\bf red zone}, if there are 10 or more breaches: under the correct model, this is expected to happen in less than 0.01\% of all cases and corresponds to $T_n \in [0.04, 1.00]$.
\end{enumerate}
Note that  we used the nominal number of breaches $(n\cdot T_n)$ for better articulation; for more details on Basel regulatory backtesting, we refer to \citepalias{Bas1996}.

While the general properties of test statistic $T_n$ have been extensively studied in the literature (see e.g. \cite{Car2009}), to the best of our knowledge the duality-based relationship that will be presented in Proposition~\ref{pr:t1.bis} has not been investigated. Before we state the result let us recall the definition of the standard empirical  VAR estimator. For a fixed confidence level $\alpha\in (0,1]$ and sample $x=(x_1,\ldots,x_n)$, where $n\in\bN$, we say that
\begin{equation}\label{eq:var.hist}
\hat\var^{\alpha}_{n}(x):=-x_{(\lfloor n\alpha\rfloor+1)},
\end{equation}
is the empirical (historical) estimator of VAR at level $\alpha\in (0,1]$, where $x_{(k)}$ is the $k$-th order statistic of the sample and $\lfloor z \rfloor$ denotes the (integer) floor of $z\in\bR$. We are now ready to present Proposition~\ref{pr:t1.bis}; see~Appendix \ref{Sec:proofs} for the proof.

\begin{proposition}\label{pr:t1.bis}
For a secured position $y=(y_1,\ldots,y_n)$ let the test statistic $T_n$ be given by \eqref{eq:Tn.first}. Then,
\begin{equation}\label{eq:t1.bis}
T_n=\inf \{\alpha\in (0,1]:\, \hat\var^{\alpha}_{n}(y) \leq 0\},
\end{equation}\
where the convention $\inf\emptyset=1$ is used.
\end{proposition}

From Proposition~\ref{pr:t1.bis} we see that to test the IM performance, the regulator can be considered as using empirical VAR estimator with different confidence levels $\alpha\in (0,1]$ and checking for which levels the position is acceptable (conservative). Noting that VAR family is monotone with respect to the target confidence level, we can find the minimal level at which we accept the position. This number is used to quantify the performance of the position. The traffic-light performance threshold $0.02$ and $0.04$ account for potential model misspecification, bias, model risks, etc. In other words, the regulator wants to ensure that the model is conservative if we (slightly) increase the confidence risk-level; cf. Remark 4.3 in \cite{PitSch2016}.

In fact, Representation~\eqref{eq:t1.bis} shows that $T_n$ could be treated as a {\it performance measure} or {\it acceptability index}; see~\cite{CheMad2009}. This family of maps is designed to assess the performance of financial positions. As the backtest statistics are meant to measure the performance of the {\it secured position}, it makes such representation very natural.

Using the duality theorem from~\cite{CheMad2009} we see that the test function $T_n$ might be seen as a map that is {\it dual} to the VAR family of maps (i.e. historical estimators $(\hat\var^{\alpha}_n)_{\alpha\in (0,1]}$) and is consequently a natural candidate for a backtest function. It should be noted that performance measures are typically defined on parameter space $\bR_{+}$ and not $(0,1]$. Nevertheless, this is only a normalisation scheme, i.e. by applying a (monotone) parameter distortion function $g(x)=\frac{1}{1+x}$ we could change the parametrisation; see Section 4.2 in~\cite{BieCiaZha2012} for more details.



 \section{Expected Shortfall backtesting}\label{S:es.bt}
In this section we define the backtesting framework for ES. The motivation behind our choice of the backtest function is driven by the dual representation behind coherent risk measures and scale-invariant acceptability indices. In a nutshell, we want to maintain relationship similar to the one outlined in Proposition~\ref{pr:t1.bis} but for ES instead of VAR. 

As before, given IM methodology, we use $y=(y_1,\ldots,y_n)$ to denote a sample of secured positions now backed by capital amounting to ES. We define test statistic $G_n$ by setting
\begin{equation}\label{eq:es.bis2}
G_n= \sum_{k=1}^{n}\frac{\1_{\{y_{(1)}+\ldots+y_{(k)}<0\}}}{n},
\end{equation}
where  $y_{(k)}$ is the $k$-th order statistic of $y$. Simply speaking, we look for the biggest number of worst realisations of the secured position that add up to a negative total, and then we divide the outcome by $n$. This framework seems to be very natural for ES -- instead of measuring the total number of exceptions we focus on the performance of the worst-case sums (or averages).

In fact, in \eqref{eq:es.bis2} we link the correct specification of the underlying ES model to its property of producing the correct worst-case conditional sums -- up to their sign. As in the VAR case, this is significantly less restrictive than imposing an overall i.i.d. assumption on the P\&L vector used in the historical ES estimator; see Section~\ref{S:preliminaries} for the discussion. 

\begin{remark}
As we are combining (summing) values of $y_{(k)}$ for different values of $k$ we need to pay particular attention to potential volatility peaks. For portfolios with highly variable risk profiles or in the presence of volatility clusters it might be advisable to apply the normalisation scheme proposed in Remark~\ref{rem:norm} and consider $\widetilde y$ instead of $y$ in \eqref{eq:es.bis2}. It should be noted that this subtlety is a result of the paradigm change coming from the VAR to ES migration rather than construction flow; see~\cite{ConDegSca2010} where the ES risk measurement statistical intricacies are discussed.
\end{remark}

\begin{remark}
A different interpretation for  \eqref{eq:es.bis2} might be achieved by splitting the secured position back into its two defining components, i.e. P\&L realisations and capital reserve estimates. It is easy to note that the nominal value $n\cdot G_n$ is equal to $k$ if and only if
\[
-\sum^k_{i=1} \textrm{P\&L}_{(i)} < \sum^k_{i=1} \hat \rho_{(i)} \quad \textrm{and}\quad  -\sum^{k+1}_{i=1} \textrm{P\&L}_{(i)} \geq \sum^{k+1}_{i=1}\hat \rho_{(i)} ,
\]
where $[\cdot]_{(i)}$ is the index correspond to $i$th order statistic of the secured position $y$. Consequently, we can say that we are summing the worst realisations of the secured position until the underlying aggregated losses are equal to their corresponding aggregated capital reserves.
\end{remark}

Before we focus on statistical aspects of \eqref{eq:es.bis2} let us show how our statistic could be embedded into the regulatory traffic-light framework. Following guidelines from \citepalias{Bas2016}, we set the reference metric to ES at level $\alpha=2.5\%$ (instead of VAR at level $\alpha=1\%$). For simplicity, and to be more aligned with the VAR framework, we refer to the nominal number of sum components $n\cdot G_n$ instead of the averages.

A correctly specified model for the 2.5\% ES reference risk metric is expected to produce negative total for up to 6 or 7 worst-case secured positions  during $n=250$ (business) days. Constructing a confidence band around that numbers an internal model (IM) is said to be in:

\begin{enumerate}[-]

\item {\bf green zone}, if the sum of the 12 smallest values of $y$ is positive: under the correct model, this is expected to happen in around 90\% of all cases and corresponds to $G_n \in[0.00, 0.05)$;\footnote{Note that the threshold value (0.05) could not be obtained by $G_n$ for $n=250$ since $0.05\times 250=12.5$ is not an integer. For better analytical traceability, we set the corresponding nominal threshold value to 12 since it is the closest (conservative) integer value.}

\item {\bf yellow zone}, if the sum of the 12 smallest values of $y$ is negative but the sum of the 25 smallest values of $y$ is positive: under the correct model, this is expected to happen in around 10\% of all cases and corresponds to $G_n\in [0.05, 0.10)$;

\item {\bf red zone}, if the sum of the 25 smallest values of  $y$ is negative: under the correct model, this is expected to happen in less than 0.01\% of all cases and corresponds to $G_n \in [0.10, 1.00]$.

\end{enumerate}
The proposed threshold levels could be modified due to regulator preference; the size of the (multiplicative) regulatory add-on could be determined by the biggest number of worst-case observations with negative sum; in the VAR case, it depended on the total number of exceptions. For completeness, we present the motivation behind our choice of the threshold levels: 

First, our thresholds levels are consistent with VAR thresholds and have a direct financial interpretation. In the VAR setting (with reference risk level $1\%$) the threshold values $2\%$ and $4\%$ could have been interpreted as the maximal acceptable risk level misspecification thresholds that were obtained by multiplying the initial confidence level by 2 and 4, respectively; see Proposition~\ref{pr:t1.bis}. Here, we do the same. In fact, we will later show that the analogue of Proposition~\ref{pr:t1.bis} is true for ES, i.e. $G_n$ is the performance measure dual to the ES risk measure family; see Proposition~\ref{pr:G1.bis}.

Second, the thresholds are chosen in such a way, that the ES risk capital levels are comparable with the corresponding VAR  risk capital levels under the normality assumption. For $X$ denoting the standard normal variable we get $\textrm{ES}^{5\%}(X)$ and $\textrm{ES}^{10\%}(X)$ equal to approximately 2.06 and 1.75, respectively. Similarly, $\textrm{VAR}^{2\%}(X)$ and $\textrm{VAR}^{4\%}(X)$ are equal to  2.05 and 1.75, respectively. Note that the regulator proposed the reference risk level change for VAR/ES migration in the similar fashion: $\textrm{ES}^{2.5\%}(X)$ is equal to 2.34, while $\textrm{VAR}^{1\%}(X)$ is equal to 2.33. This allows a smooth transition between VAR and ES frameworks (at least if the secured position distribution is close to normal). 

Third, we show that proposed thresholds lead to statistical framework that is aligned with VAR backtest. Even under extreme specification imposed on the null distribution the proposed thresholds lead to statistical confidence thresholds close to 95\% and 99.99\%.  Consequently, the thresholds could be considered as (almost) model-independent; see Section~\ref{S:test.performance} for details.

Fourth, the numerical simulations shows that the proposed threshold choice leads to a framework that is aligned with the old VAR framework; the detailed explanation of this fact will be given in Section~\ref{S:test.performance} and Section~\ref{Sec:empirical}.

Now, as stated in the beginning of this section, we want to show that the test statistic $G_n$ is a performance measure that is dual to the ES family of empirical estimators. For a fixed confidence level $\alpha\in (0,1]$ and sample $x=(x_1,\ldots,x_n)$, where $n\in\bN$, we call
\begin{equation}\label{eq:es.hist}
\hat{\textrm{ES}}^{\alpha}_{n}(x):=-\left(\frac{ \sum_{i=1}^{n}x_i\1_{\{x_i+\hat{\var}^{\alpha}_n(x)\leq 0\}}}{\sum_{i=1}^{n}\1_{\{x_i+\hat{\var}^{\alpha}_n(x)\leq0\}}} \right),
\end{equation}
the empirical (historical) estimator of ES at level $\alpha\in (0,1]$, where the map $\hat{\var}^{\alpha}_n(\cdot)$ is defined in~\eqref{eq:var.hist}. We are now ready to present Proposition~\ref{pr:t1.bis}; see~Appendix \ref{Sec:proofs} for the proof.

\begin{proposition}\label{pr:G1.bis}
For a secured position $y=(y_1,\ldots,y_n)$ let the test statistic $G_n$ be given by \eqref{eq:es.bis2}. Then,
\begin{equation}\label{eq:es.bis}
G_n=\inf \{\alpha\in (0,1):\, \hat{\textnormal{ES}}^{\alpha}_{n}(y) \leq 0\}.
\end{equation}
where the convention $\inf\emptyset=1$ is used
\end{proposition}

As in the VAR case, we look for the minimal confidence level which makes the secured position acceptable. The intuition behind representation~\eqref{eq:es.bis} is similar as in the VAR case and is omitted here. Let us recall that from~\cite{CheMad2009} we know that $G_n$ is a performance measure designed to quantify the performance of $y$ under ES framework, and is obtained using the dual representation between coherent risk measures and scale-invariant acceptability indices; see Section 4.2 in~\cite{BieCiaZha2012} for details.

\section{Theoretical test statistic distribution}\label{S:test.performance}
In this section we want to check how the distribution of the ES test statistic $G_n$ looks like for $n=250$ and how it is related to the corresponding VAR statistic $T_n$. For transparency, and to be more aligned with the regulatory framework, we use nominal values $n\cdot G_n$ and $n\cdot T_n$ instead of $G_n$ and $T_n$, respectively.

For selected {\it a priori} distributions, we simulate a strong Monte Carlo (MC) sample of size 50\,000, where each strong sample corresponds to 250 observations (portfolio P\&Ls). To construct the corresponding VAR 1\% and ES 2.5\% secured positions we add the true risk values to the observations. Then, for each MC run we calculate the value of the nominal statistics.

\begin{figure}[htp!]
\begin{center}
\includegraphics[width=7cm]{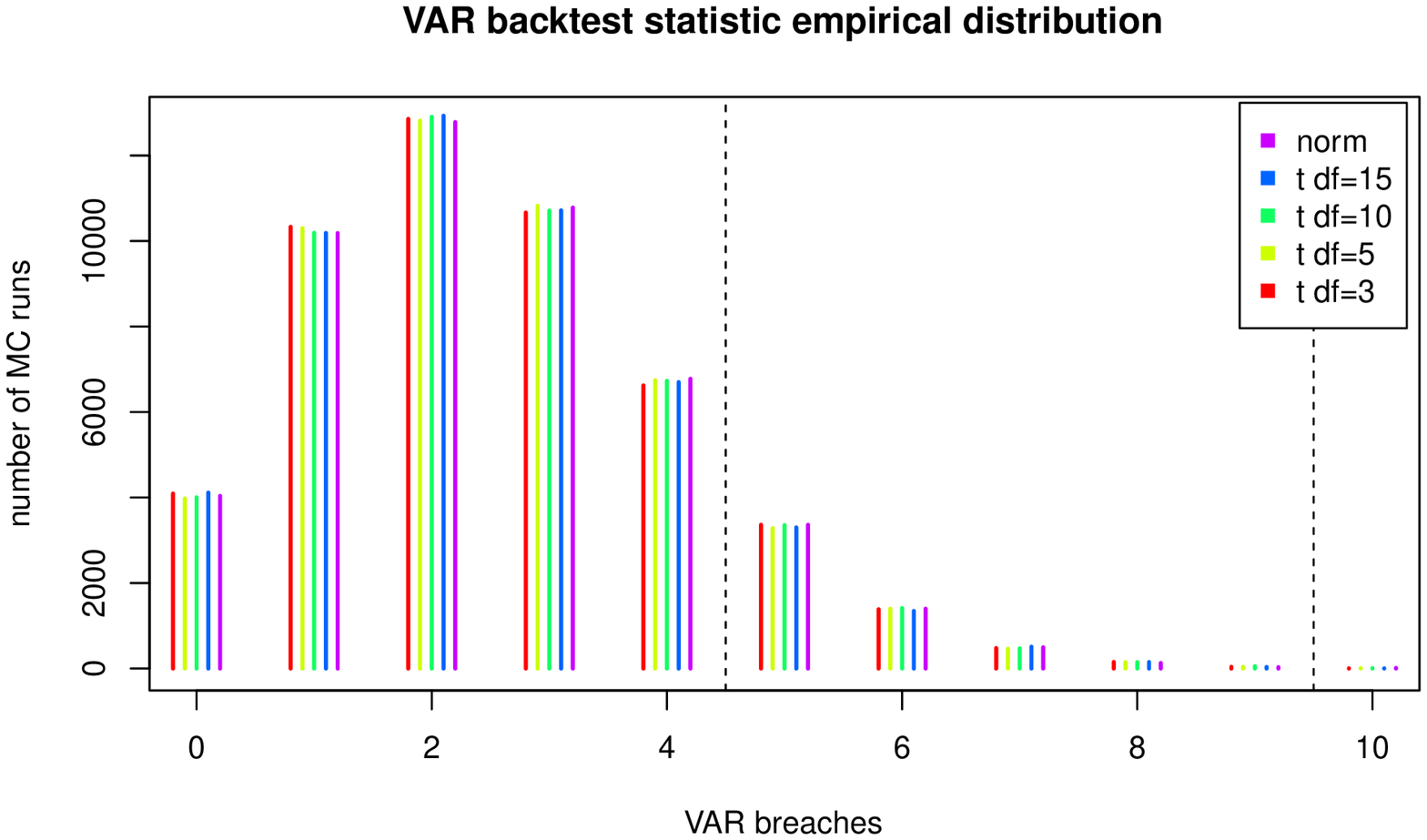}
\includegraphics[width=7cm]{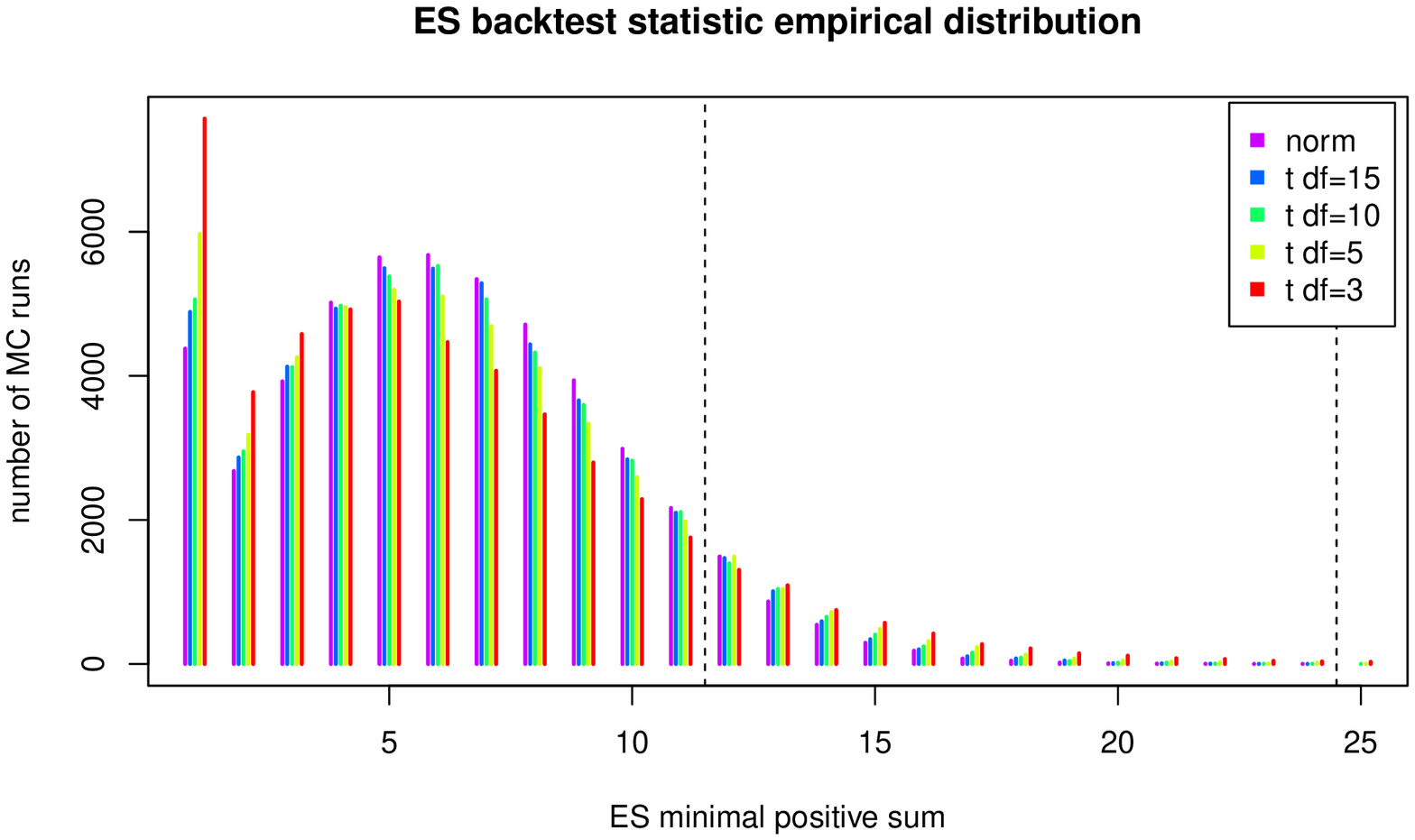}
\end{center}
\caption{We present empirical probability mass functions of the nominal VAR backtest statistics $n\cdot T_n$ (left) and nominal ES backtest statistic $n\cdot G_n$ (right) for $n=250$. We consider five different a priori distributions, and construct the secured samples using true risk capital reserve add-ons. Note that the probability mass function for VAR corresponds to Bernoulli probability mass function with $p=0.99$. We can see that ES backtest statistic is remarkably stable under extreme conditions, i.e under t-student distribution with $\nu=3$ degrees of freedom. Dotted lines indicate the proposed traffic-light thresholds. The values were obtained using a 50\,000 Monte Carlo run. }
\label{F:stat1}
\end{figure}

We take the standard normal distribution as well as t-student distribution with $3$, $5$, $10$, and $15$ degrees of freedom. 
The obtained empirical probability mass functions of nominal test statistics are presented in Figure~\ref{F:stat1}, while the cumulative empirical distribution values nearby the traffic-light threshold could be found in Table~\ref{T:stat1}.

For VAR, the simulated distribution correspond to the Bernoulli distribution with probability of success parameter equal to $0.99$. One can see that for nominal $T_n$ the cumulative probabilities for the thresholds oscillates around $0.95$ and $0.99$, respectively. For nominal $G_n$, the cumulative probabilities depend on the underlying distribution but are remarkably stable nearby the thresholds. The difference of cumulative probabilities for limit cases for the first threshold is around $0.03$ which suggest that the test statistic is robust with respect to the underlying distribution choice. As expected, the red-zone threshold is very close to 1 in all cases.

We have also tested the distribution of $G_n$ under various GARCH(1,1) model specifications imposed by market data. The results are very similar to the ones presented in Figure~\ref{F:stat1} and are omitted for brevity; they are available from the authors upon request.

Finally, please recall that in order to remove volatility clusters or risk-profile changes that might result in the non i.i.d. behaviour of the secured sample, one might want to consider applying the normalisation scheme outlined in Remark~\ref{rem:norm}.

\begin{table}[htp!]
\centering
\begin{tabular}{|r|r|rr|rr|rr|rr|}
  \hline
\multicolumn{2}{|c|}{Risk metric} & \multicolumn{4}{c|}{VAR} & \multicolumn{4}{c|}{ES} \\ 
  \hline
\multicolumn{2}{|c|}{Statistic value} & 4 & 5 & 9 & 10 & 11 & 12 & 24 & 25 \\ 
\hline
 t-student& $\nu=3\phantom{0}$ &0.8914 & 0.9586 & 0.9998 & 0.9999 & 0.8944 & 0.9205 & 0.9967 & 0.9973 \\ 
  &$\nu=5\phantom{0}$ &  0.8931 & 0.9588 & 0.9998 & 1.0000 & 0.9074 & 0.9372 & 0.9998 & 0.9999 \\ 
  &$\nu=10$ &0.8909 & 0.9580 & 0.9998 & 1.0000 & 0.9185 & 0.9464 & 1.0000 & 1.0000 \\ 
 & $\nu=15$ &  0.8930 & 0.9590 & 0.9999 & 1.0000 & 0.9224 & 0.9518 & 1.0000 & 1.0000 \\ 
 \hline
\multicolumn{2}{|c|}{normal}& 0.8913 & 0.9585 & 0.9997 & 1.0000 & 0.9292 & 0.9591 & 1.0000 & 1.0000 \\ 
   \hline
\end{tabular}
\caption{The table presents the cumulative (empirical) distribution values of the nominal backtesting statistics for VAR and ES, for large samples from various pre-defined distributions. The distribution of VAR test statistics correspond to Bernoulli distribution with $p=0.99$, and the theoretical threshold values are 0.9588, and 0.9999 (for 5 and 10 exceeds, respectively). One can see that ES backtesting statistic is stable even in extreme conditions (for $\nu=3$) and the cumulative probability for the thresholds is comparable to VAR (for values 12 and 25). The values were obtained using a 50\,000 Monte Carlo run.}
\label{T:stat1}
\end{table}

\section{Empirical study}\label{Sec:empirical}
In this small empirical study we assess the performance of ES backtesting framework using various sets of real market and simulated data. In particular, we show that the proposed framework allows a smooth transition from VAR backtesting into ES backtesting. Also, we check the performance of our testing framework in reference to 'Test 2' from~\cite{Acerbi2014Risk}.

For transparency, we decided to use only normal and empirical estimators both for VAR and ES. We pick those two estimators because of their different specification: the first one is a parametric estimator that will allow us to investigate the behaviour of the misspecified methodology (e.g. if we pick a sample from t-student instead of normal), while the second one is a non-parametric estimator that should be model independent.

We keep the estimation learning period fixed and equal to one year (250 observations). As in the standard VAR framework, the backtesting period is equal to 1-day, and the rolling window length is set to 250. Thus, for a single test run we need 500 consequent observations to compare the estimated capital reserve with realised P\&L vectors and to perform the backtesting exercise. Starting from day 251 for each day we sum the estimated capital reserve and the realised P\&L to get the realised secured position value. We do that up to day 500 in order to obtain secured sample $y$ of size 250.

Let $x=(x_1,\ldots,x_{500})$ denote the vector of 500 consequent realised 1-day P\&L vectors. For $i=1,\ldots,250$\,, the $i$-th backtesting day VAR estimated capital reserves are equal to
\begin{align*}
\hat\var^{\textrm{norm}}_{i}(x) &:=-\left(\bar\mu_i +\bar\sigma_i\Phi^{-1}(0.01)\right),\\
\hat\var^{\textrm{emp}}_{i}(x) &:=\hat\var^{0.01}_{250}(x_i,\ldots,x_{i+249}),
\end{align*}
where the empirical estimator $\hat\var^{0.01}_{250}$ is defined in~\eqref{eq:var.hist},  $\bar\mu_i$ is the $i$-th backtesting day efficient mean estimator, and $\bar\sigma_i$ is the $i$-th backtesting day efficient standard deviation estimator, i.e. 
\[
\bar \mu_i= \frac{1}{250}\sum^{249}_{k=0}x_{i+k},\quad \textrm{ and }\quad \bar \sigma_i = \sqrt{\frac{1}{249}\sum^{249}_{k=0}\Big(x_{i+k}- \bar\mu_i\Big)^2}.
\]
Similarly, the $i$-th backtesting day ES estimated capital reserves are given by
\begin{align*}
\hat{\textrm{ES}}_{i}^{\textrm{norm}}(x) & :=-\left(\bar\mu_i+\bar{\sigma}_i\frac{\phi(\Phi^{-1}(0.025))}{1-0.025}\right),\\
\hat{\textrm{ES}}_{i}^{\textrm{emp}}(x) & :=\hat{\textrm{ES}}^{0.025}_{250}(x_i,\ldots,x_{i+249}),
\end{align*}
where the empirical estimator $\hat{\textrm{ES}}^{0.025}_{250}$ is given in~\eqref{eq:es.hist}, $\phi$ relates to the density function of the standard normal, and $\Phi$ is the distribution function of the standard normal; see~\cite{PitSch2016} for details. Next, we construct secured position vectors for all estimated capital reserves following the logic from \eqref{eq:secures.position}, i.e. for $i=1,\ldots,250$ we set
\begin{align}
y^{\var^{\textrm{norm}}}_i & :=x_{i+250} +\hat\var^{\textrm{norm}}_{i}(x), & 
y^{\textrm{ES}^{\textrm{emp\phantom{h}}}}_i & :=x_{i+250} +\hat{\textrm{ES}}_{i}^{\textrm{emp}}(x),\nonumber\\ 
y^{\var^{\textrm{emp\phantom{h}}}}_i & :=x_{i+250} +\hat\var^{\textrm{emp}}_{i}(x), & y^{\textrm{ES}^{\textrm{norm}}}_i & :=x_{i+250} +\hat{\textrm{ES}}_{i}^{\textrm{norm}}(x).\label{eq:secured.ea}
\end{align}

For simplicity, instead of the realised P\&Ls we compare the realised return rates; cf. Remark~\ref{rem:norm}. Note that because both VAR and ES are positively homogeneous risk measures, the resulting analysis produce consistent outputs. Also, as in Section~\ref{S:test.performance}, instead of presenting the results for $T_n$ and $G_n$ we consider the nominal values, i.e. number of exceptions ($n\cdot T_n$) and biggest number of worst-case scenarios which sum up to negative total ($n\cdot G_n$).


To test our framework we conduct two tests. In Section~\ref{S:consistency.ea}, we investigate the alignment between the ES backtest and the standard regulatory VAR backtest. Using market and simulated data we compare the outcomes of both backtests and check their consistency. In Section~\ref{S:relation}, we compare our backtest with 'Test 2' proposed in~\cite{Acerbi2014Risk}.

\subsection{Consistency between VAR and ES backtesting frameworks}\label{S:consistency.ea}
In this section we perform backtests described in Section~\ref{S:var.bt} (for historical and normal VAR estimators) and Section~\ref{S:es.bt} (for historical and empirical ES estimators) for various data sets. We classify the outcomes into green, yellow, or red zone by computing the nominal values of $T_n$ and $G_n$, and by applying the classifications scheme outlined in Table~\ref{T:classification}.

To compare VAR and ES backtesting frameworks we check the numbers of common classifiers for normal estimators and for the empirical estimators. For completeness, we also present the plots that confronts nominal values of $T_n$ with the nominal values of $G_n$ for all considered samples.
\begin{table}[htp!]
\centering
\begin{tabular}{|r|r|r|}
\hline
Zone & VAR   & ES \\ 
(color) & (number of exceptions) & (worst-case scenarios with negative sum) \\ 
\hline
Green& 0--4 &  0--11 \\
 Yellow&   5--9 &  12--24 \\ 
 Red&    10+ &   25+ \\ 
 \hline
\end{tabular}
\caption{Backtest threshold values for VAR and ES backtesting exercises for a single dataset. For VAR, the threshold should be confronted with the total number of exceptions that occurred in the sample ($n\cdot T_n$), i.e days $i$ for which we have $y_i<0$. For ES, the value should be confronted with the biggest number of days for which the worst-case scenario sum up to negative total $(n\cdot G_n$), i.e. maximal $m\in\bN$ such that $y_{(1)}+\ldots+y_{(m)}<0$, where $y_{(k)}$ is the $k$-th order statistic of $y$.}
\label{T:classification}
\end{table}

For transparency, we split the analysis into two parts: in Section~\ref{sec:FFdata} we consider market data while in Section~\ref{sec:simulated.test} we focus on the simulated data.

\subsubsection{Fama \& French library data}\label{sec:FFdata}
\begin{figure}[htp!]
\begin{center}
\includegraphics[width=14cm]{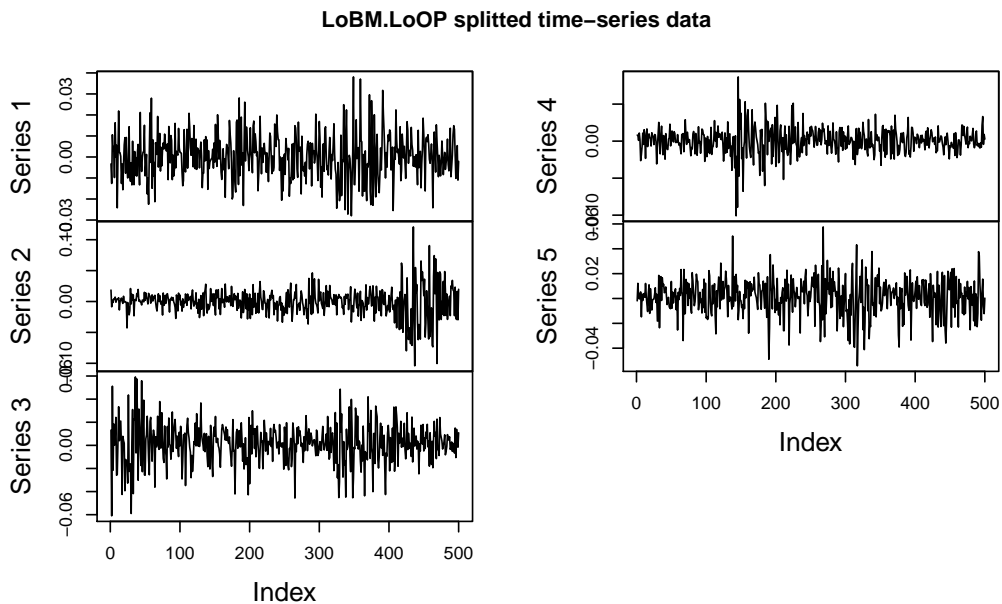}
\end{center}
\caption{Return rate time series for 5 different subsets for the first portfolio from Fama \& Fench data library.}\label{F:FF}
\end{figure}

We use daily returns from the data library \cite{FamFre2015}. We take returns of 25 portfolios formed on book-to-market and operating profitability in the period from 27.01.2005 to 01.01.2015 ($2500$ observations). For each portfolio, we split the data into disjoint subsets of 500 consecutive days, and we obtain exactly 125 different $x$-samples. Five different time-series corresponding to the first portfolio are presented in Figure~\ref{F:FF}. One can see that the time-series are non i.i.d. and exhibit volatility clustering effects, so that the backtesting classification should not be too uniform. In other words, we should get many yellow-zone and red-zone classifications, both for VAR and for ES.

\begin{figure}[htp!]
\begin{center}
\includegraphics[width=5cm]{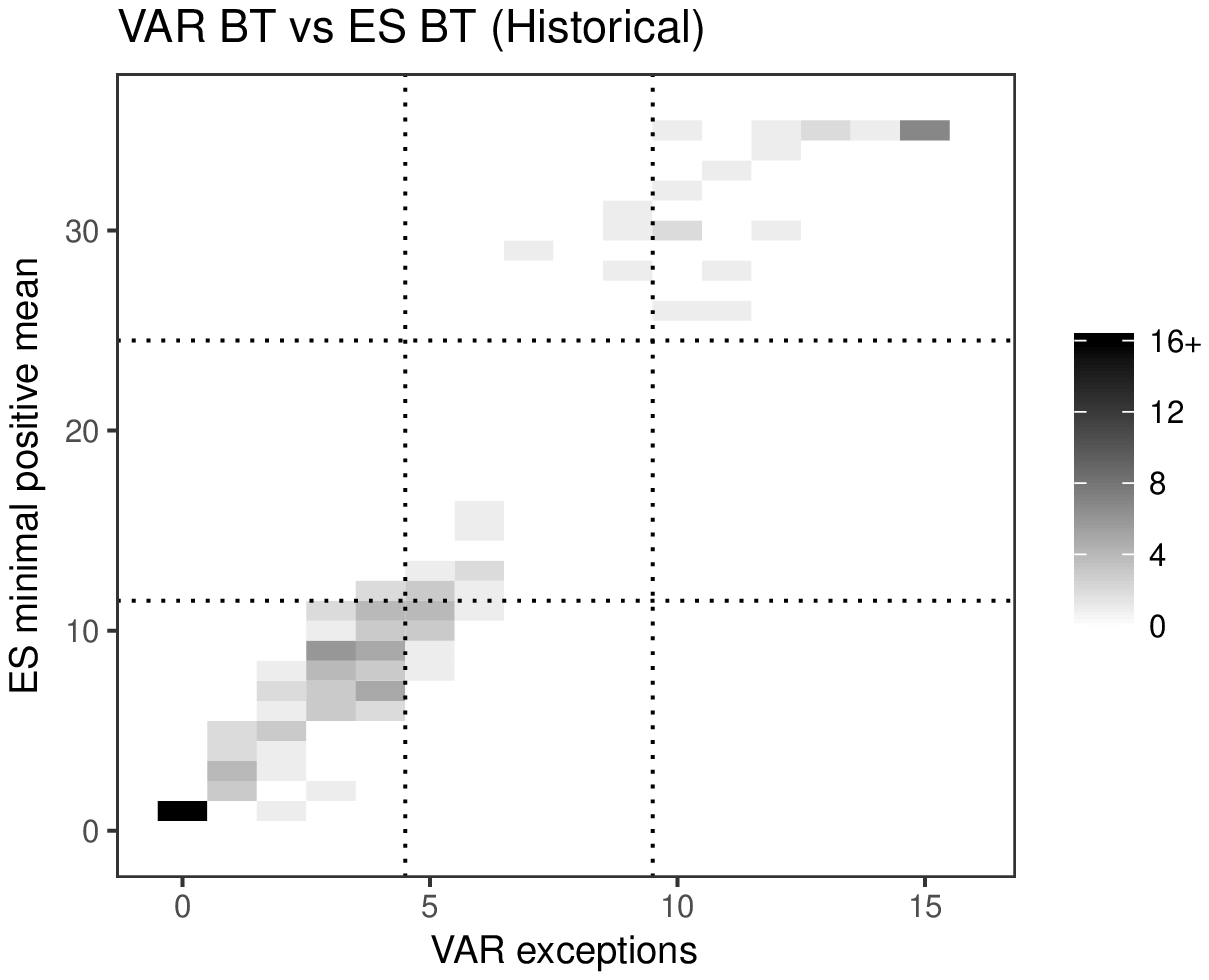}
\includegraphics[width=5cm]{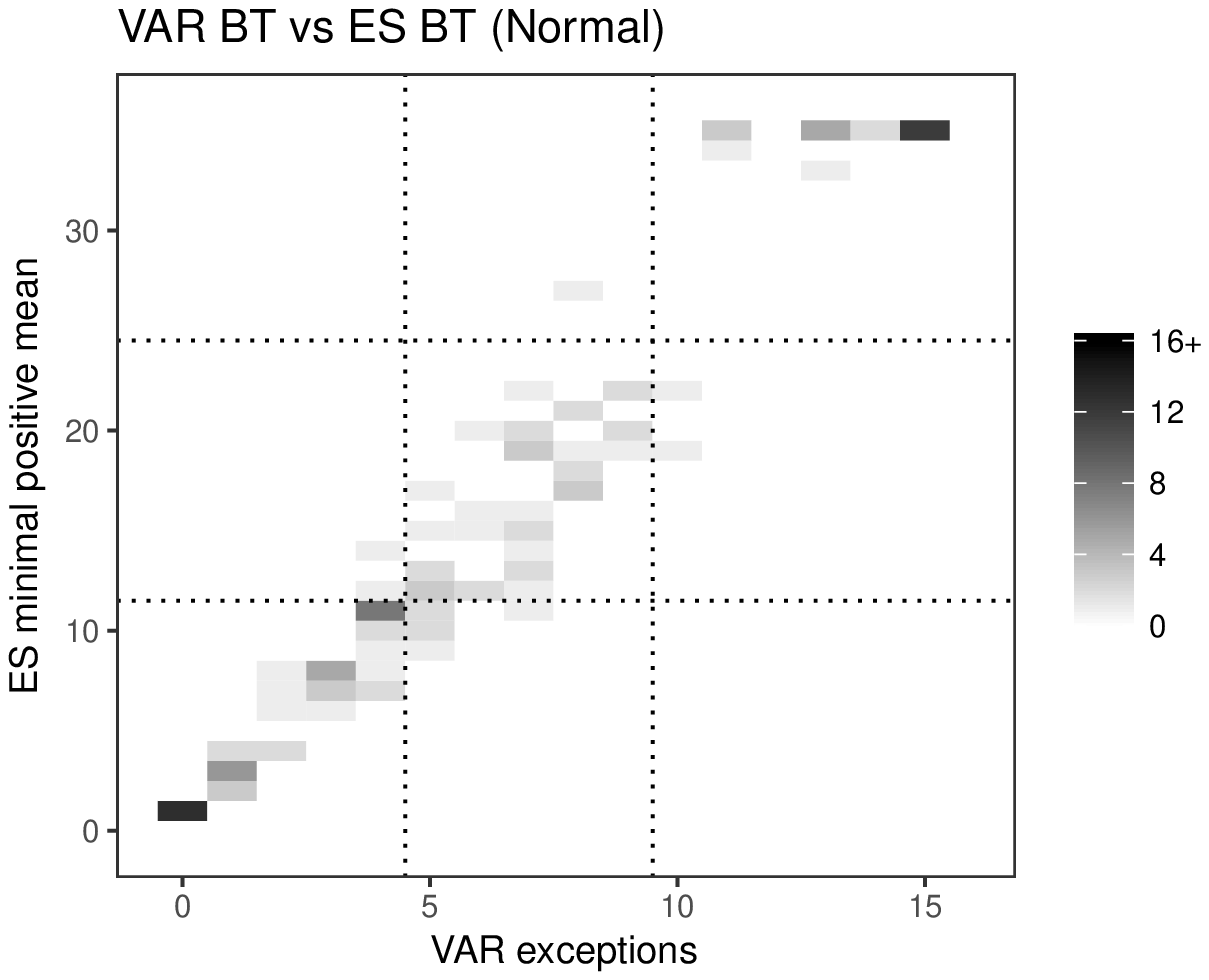}
\end{center}
\caption{Backtesting results for Market Data. For each sample $x$ we perform both VAR backtesting and ES backtesting for historical (left) and normal (right) estimators. We report number of VAR exceptions ($n\cdot T_n$) as well as biggest number of worst-case outcomes with the total negative sum ($n\cdot G_n$). The resulting 125 points are presented on the heatmap. The darker the colour, the more samples were assigned to a given cluster; see legend on the right.}\label{F:market}
\end{figure}

We perform backtesting exercise 125 times for all four estimators and classify the outcomes. The aggregated results for normal and empirical estimators are presented in Table~\ref{T:market}. See Figure~\ref{F:market} for more detailed comparative plots. For brevity, we have truncated the outcomes: if there were more than 15 VAR breaches we limit the output of $n\cdot T_n$ to 15 (in that case the model is classified clearly into red-zone). We impose a similar upper bound on $n\cdot G_n$ that is equal to 35.

\begin{table}[htp!]
\centering
\begin{tabular}{rc|rrr}
&R &   0 &   4 &  21 \\ 
$\textrm{ES}^{\textrm{hist}}$ & Y &   2&  9 &   0 \\ 
&G &  79 &  10 &   0 \\
\cline{2-5}
& & G & Y & R \\ 
\multicolumn{2}{c}{}  &\multicolumn{3}{c}{$\var^{\textrm{hist}}$}   \\ 
\end{tabular}
\qquad\qquad
\begin{tabular}{rc|rrr}
  &R &    0 &   1 &  24 \\ 
$\textrm{ES}^{\textrm{norm}}$& Y &    2 &  38 &   2 \\ 
&G &  52 &  6 &   0 \\
 \cline{2-5}
& & G & Y & R \\ 
\multicolumn{2}{c}{}  &\multicolumn{3}{c}{$\var^{\textrm{norm}}$ }   \\ 
\end{tabular}
\caption{The tables shows results for market data, for 125 different backtesting exercises for VAR and ES. One can see that the classification for VAR backtesting framework and ES backtesting framework is consistent, i.e. the values in the diagonal are higher compared to the numbers outside the diagonal).}
\label{T:market}
\end{table}

The results show good consistency between VAR and ES backtesting frameworks. As expected, there are more yellow and red zone outputs for normal estimators in both cases due to the fact that the data is non-normal.

\subsubsection{Simulated data}\label{sec:simulated.test}
In this section we repeat the testing outlined in Section~\ref{sec:FFdata} for the simulated data from normal, skew t-student, and GARCH(1,1) models. For each of the 125 market time series data from Section~\ref{sec:FFdata} and each parametric model, we fit model parameters and simulate the appropriate random sample. We have decided to take eight independent picks for each fit, and thus instead of 125 new realisations we have 1000 realisations. We repeat all the steps from the previous section. For brevity we do not present the analogues of Table~\ref{T:market} and focus on the comparative plots.

\begin{figure}[htp!]
\begin{center}
\includegraphics[width=5cm]{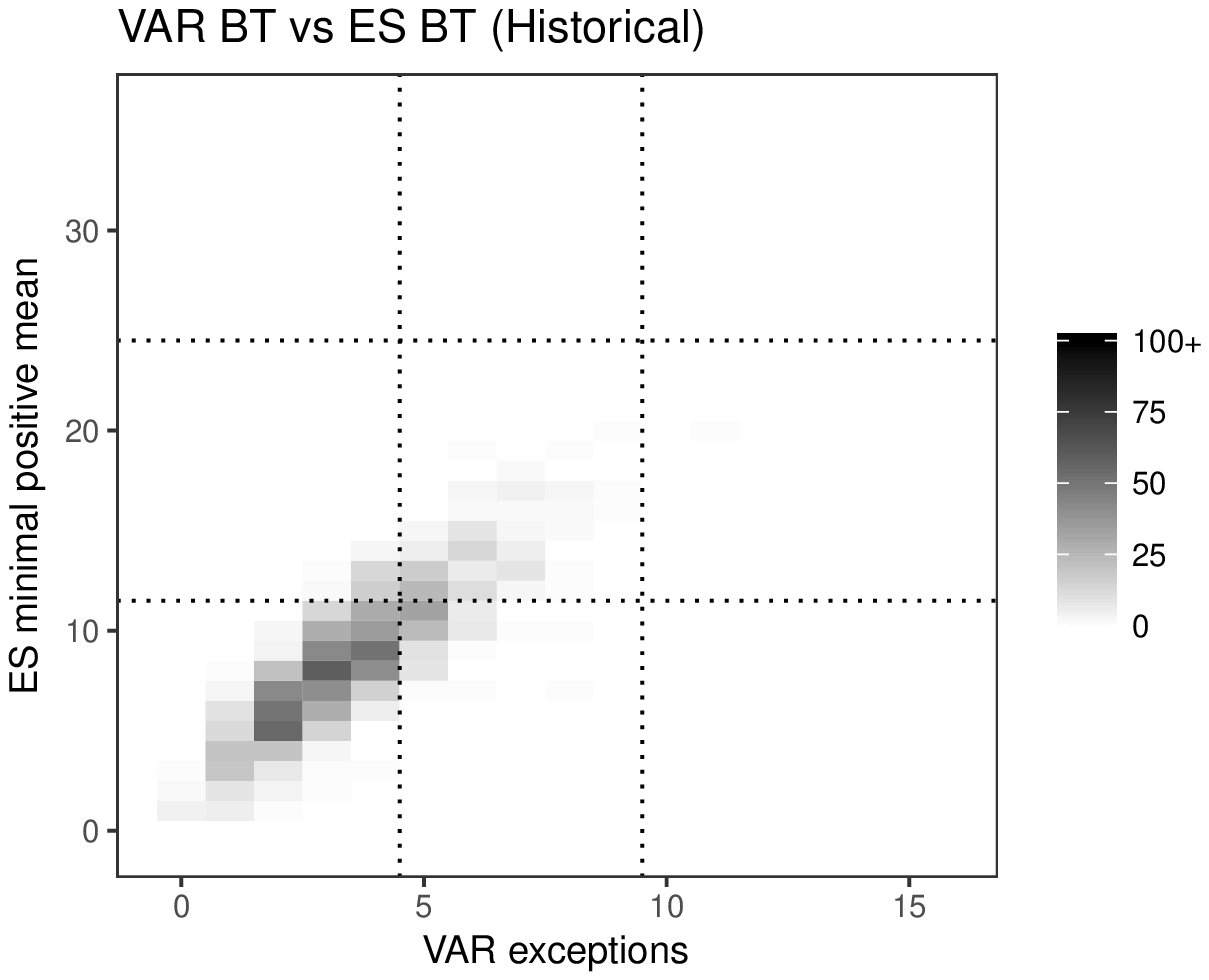}
\includegraphics[width=5cm]{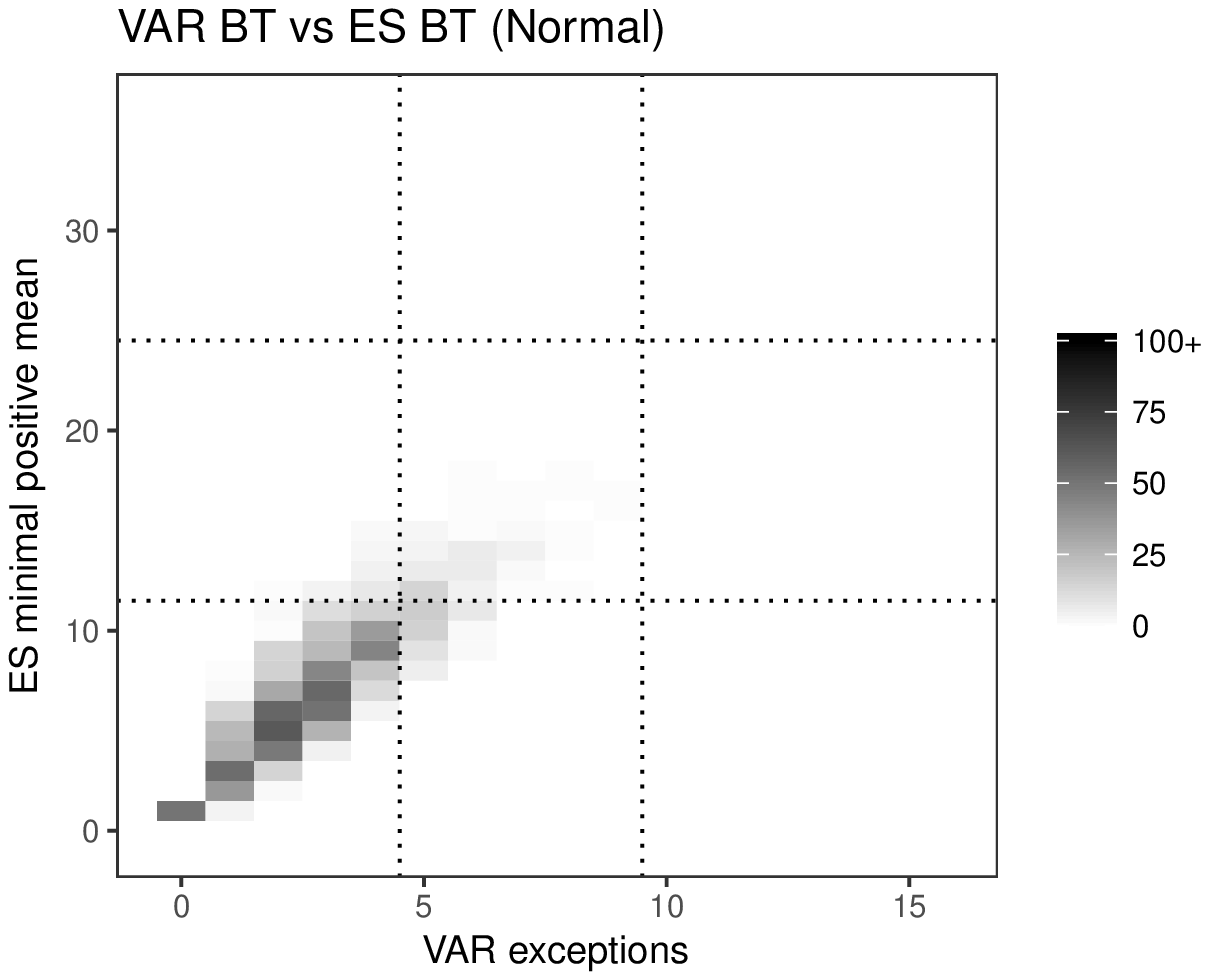}
\end{center}
\caption{Backtesting results for Normal simulated data; see Figure~\ref{F:market} caption for details.}\label{F:normal}
\end{figure}

\begin{figure}[htp!]
\begin{center}
\includegraphics[width=5cm]{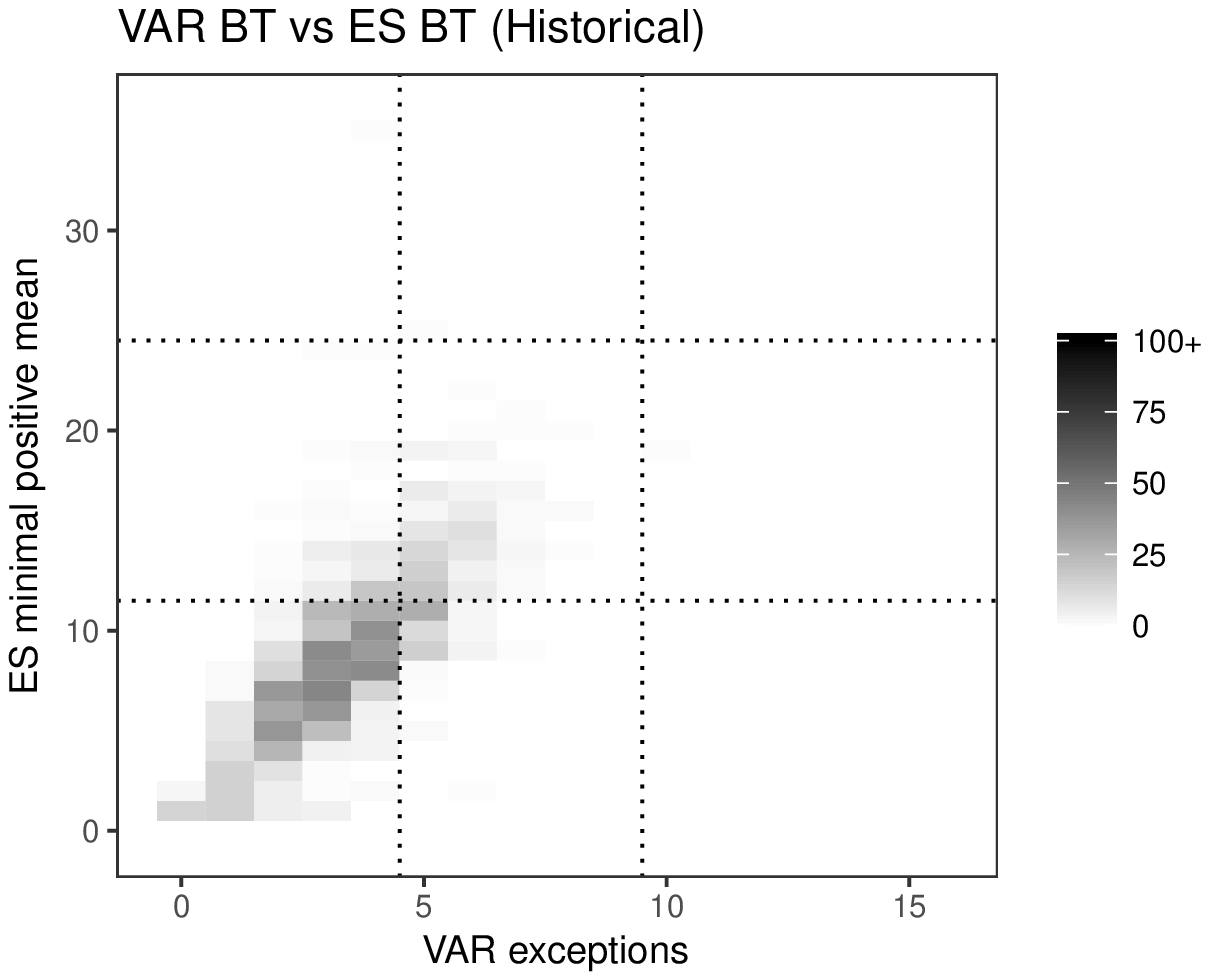}
\includegraphics[width=5cm]{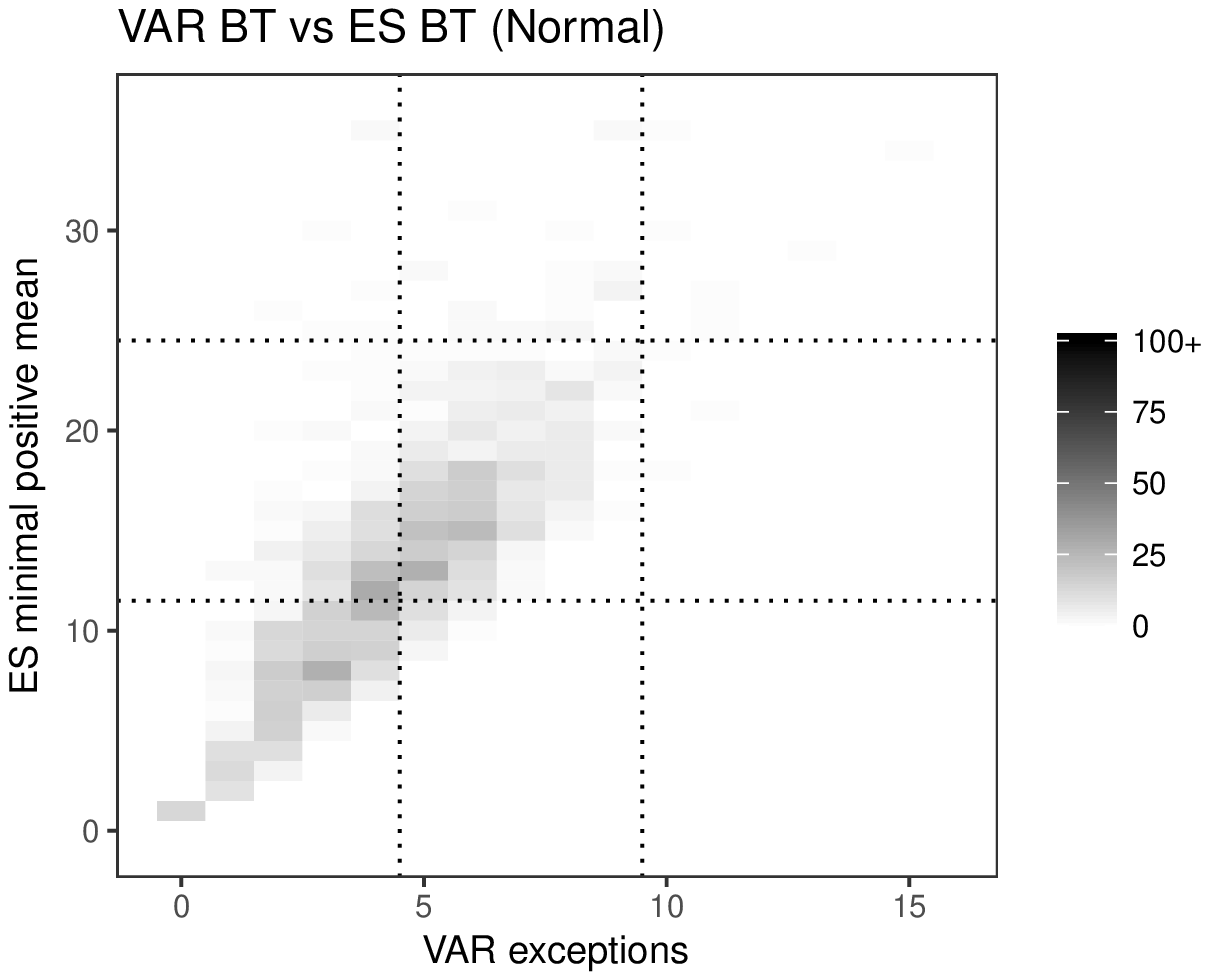}
\end{center}
\caption{Backtesting results for skew t-Student simulated data; see Figure~\ref{F:market} caption for details.}\label{F:SkewStud}
\end{figure}

\begin{figure}[htp!]
\begin{center}
\includegraphics[width=5cm]{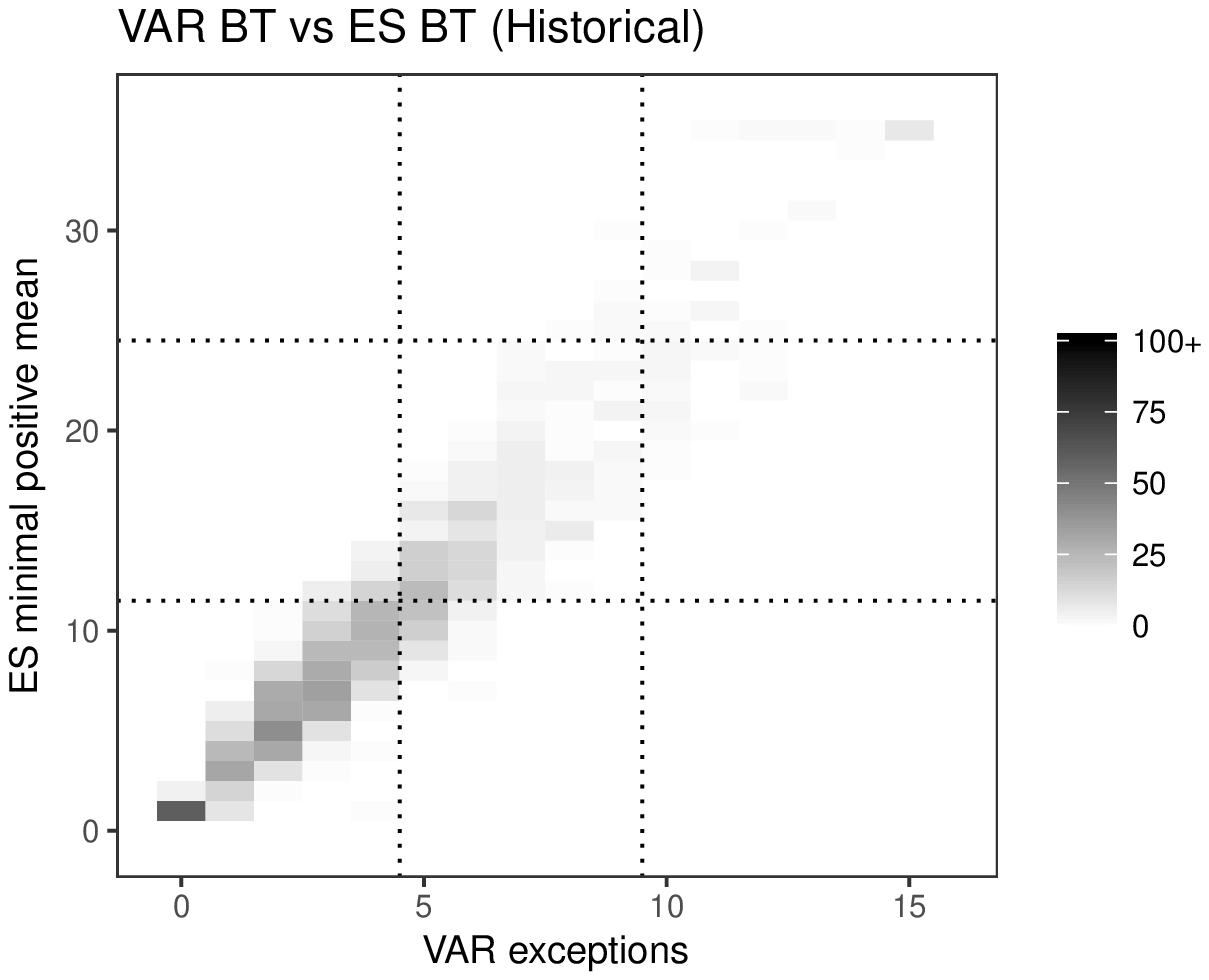}
\includegraphics[width=5cm]{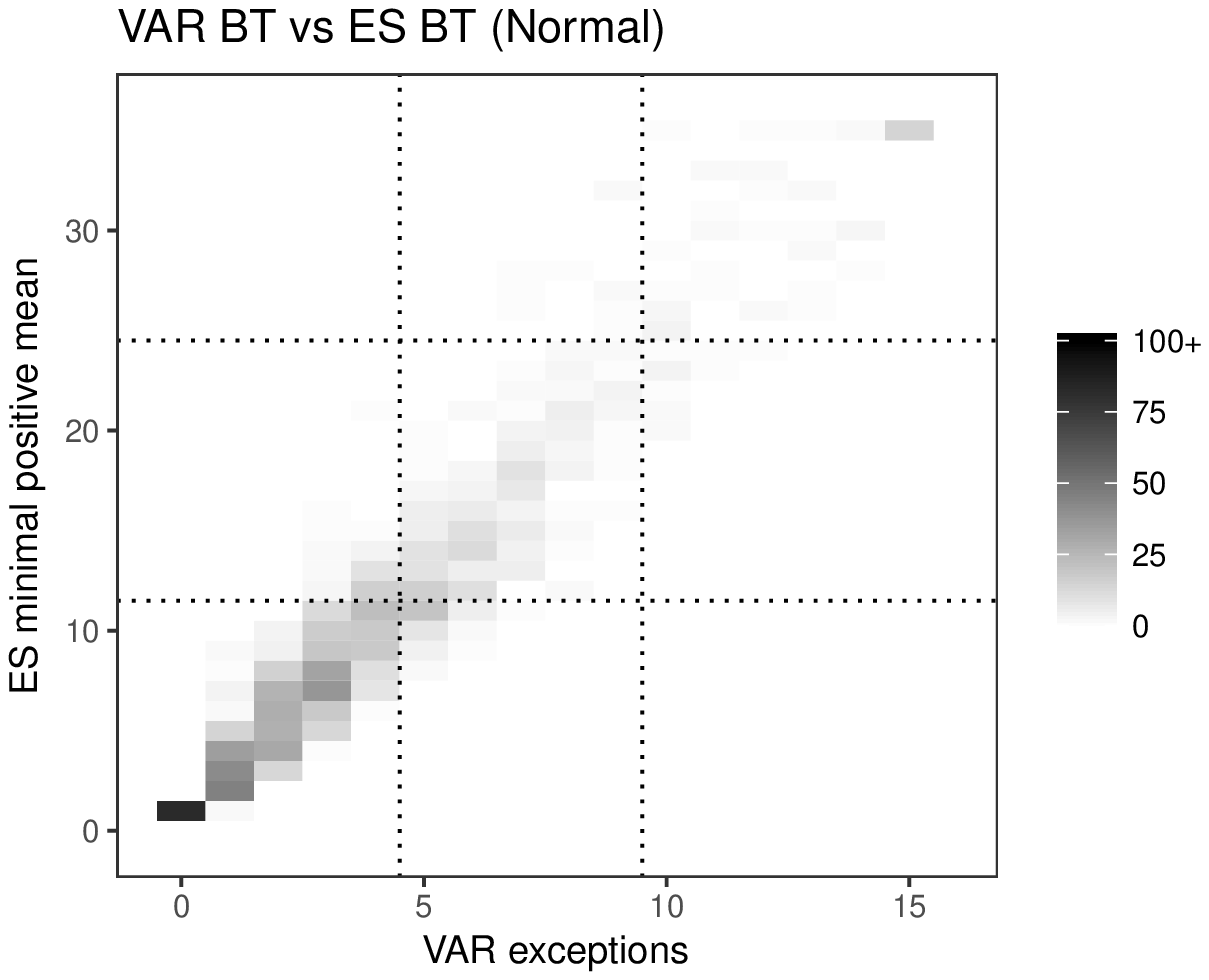}
\end{center}
\caption{Backtesting results for GARCH(1,1) simulated data with normal innovations; see Figure~\ref{F:market} caption for details.}\label{F:Garch}
\end{figure}

\begin{figure}[htp!]
\begin{center}
\includegraphics[width=5cm]{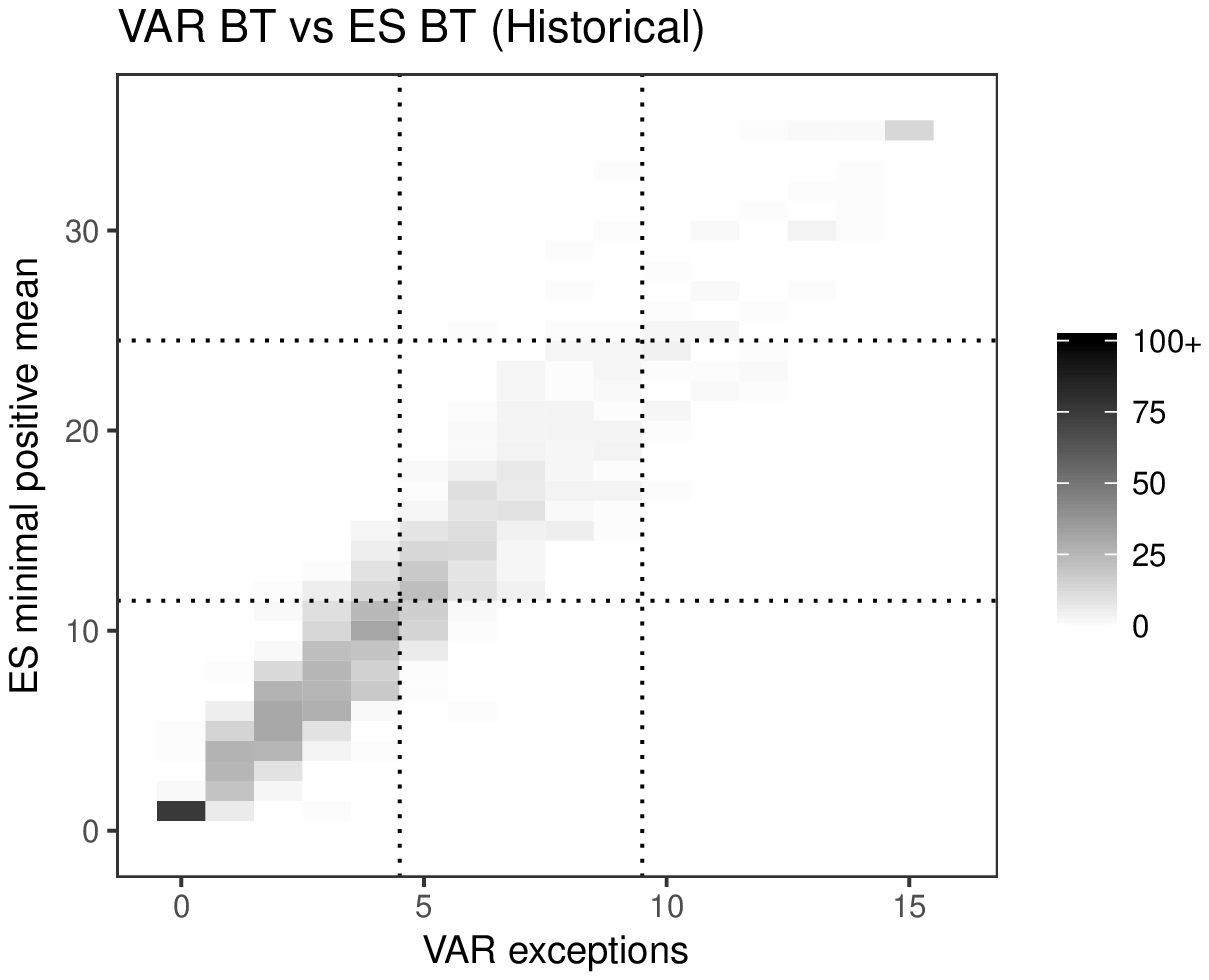}
\includegraphics[width=5cm]{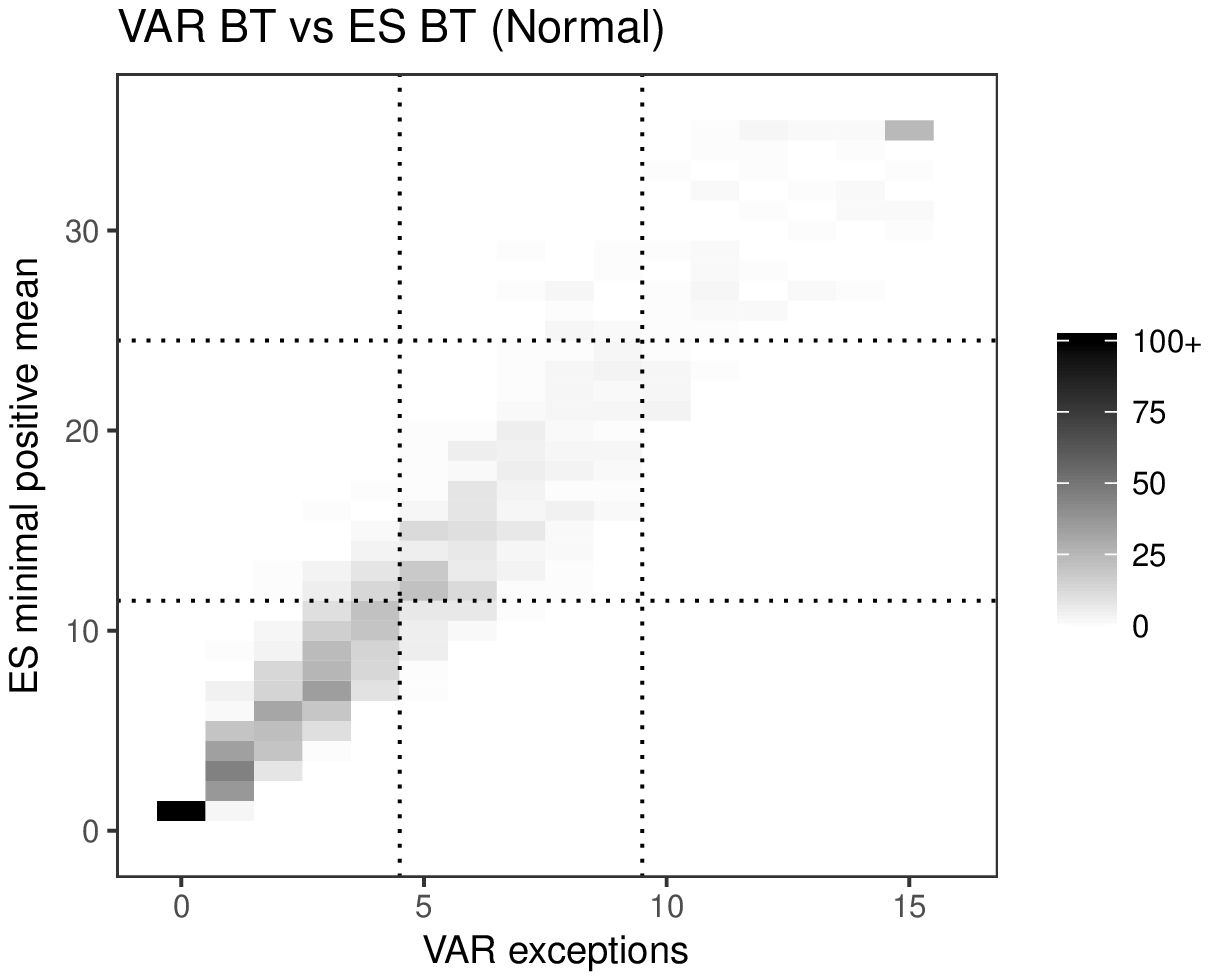}
\end{center}

\caption{Backtesting results for GARCH(1,1) simulated data with skewed t-student innovations; see Figure~\ref{F:market} caption for details.}\label{F:Garch1}
\end{figure}

The results for normal data are presented in Figure \ref{F:normal}; the results for skewed t-student are presented in Figure~\ref{F:SkewStud}; the results for GARCH(1,1) with normal innovations are presented in Figure~\ref{F:Garch}; the results for GARCH(1,1) with skew t-student innovations are presented in Figure~\ref{F:Garch1}.

One can see that the results are satisfactory. While for the normal data we get similar performance, for the t-student case there are differences. As expected, normal estimators is not performing well because it fails to predict the correct behaviour in the tail. Note that this feature is better visible in ES backtesting framework, because of the construction of the estimator. Also, note that for the GARCH data, for which the standard i.i.d. assumption is not satisfied, both backtesting frameworks classify many outputs into yellow or red zone. Also, the ES backtest better captures heavy tailed behaviour which is reflected in more conservative results for normal estimator case. Overall, our framework gives results which are consistent the standard VAR backtest.

\subsection{Relation to 'Test 2' }\label{S:relation}
In this section we show the relation between our framework and ES backtesting framework described as 'Test 2' in~\cite{Acerbi2014Risk}. For consistency, we present results for the market data and the simulated normal data from the previous testing exercise; see Section~\ref{sec:FFdata} and Section~\ref{sec:simulated.test} for details. For simplicity, we only consider results for the the historical (empirical) estimator. Results for other datasets and the normal estimators are similar and are available upon request.

Before showing the results, let us briefly outline the construction of the test statistic referred to as {\it Test 2} in~\cite{Acerbi2014Risk}. As before, we assume that we have sample $x=(x_1,\ldots,x_{500})$ and IM approach for both VAR and ES at level $\alpha\in (0,1)$. As in the previous case, for $i$-th backtesting day (where $i=1,2,\ldots,250$) we estimate capital reserves $\hat{\var}^{\textrm{norm}}_i(x)$ and $\hat{\textrm{ES}}^{\textrm{norm}}_i(x)$ using learning period data $(x_i,\ldots, x_{i+249})$ and confront the estimators with realised value $x_{i+250}$. In our setting, the test statistic given in Equation (6) in \cite{Acerbi2014Risk} is given by
\begin{align}\label{Z}
Z :=\left(\frac{1}{250}\sum_{i=1}^{250}\frac{x_{i+250}\, \ind{x_{i+250}+\hat \var^{\textrm{norm}}_{i}(x)<0 }}{\alpha \,\hat{\textrm{ES}}^{\textrm{norm}}_{i}(x)}\right)+1.
\end{align}
Under the null-hypothesis (stating that the IM is conservative) we get that the value of $Z$ is non-positive. The alternative hypothesis  corresponding to risk underestimation is reflected in positive values of the test statistic $Z$ (note that we use different sign convention). Following authors' suggestion we have set values 0.7 and 1.8 as the test-statistic threshold values for the test and get a three-zone classification scheme; see~\cite{Acerbi2014Risk} for detailed description of the testing framework.

\begin{figure}[htp!]
\begin{center}
\includegraphics[width=5cm]{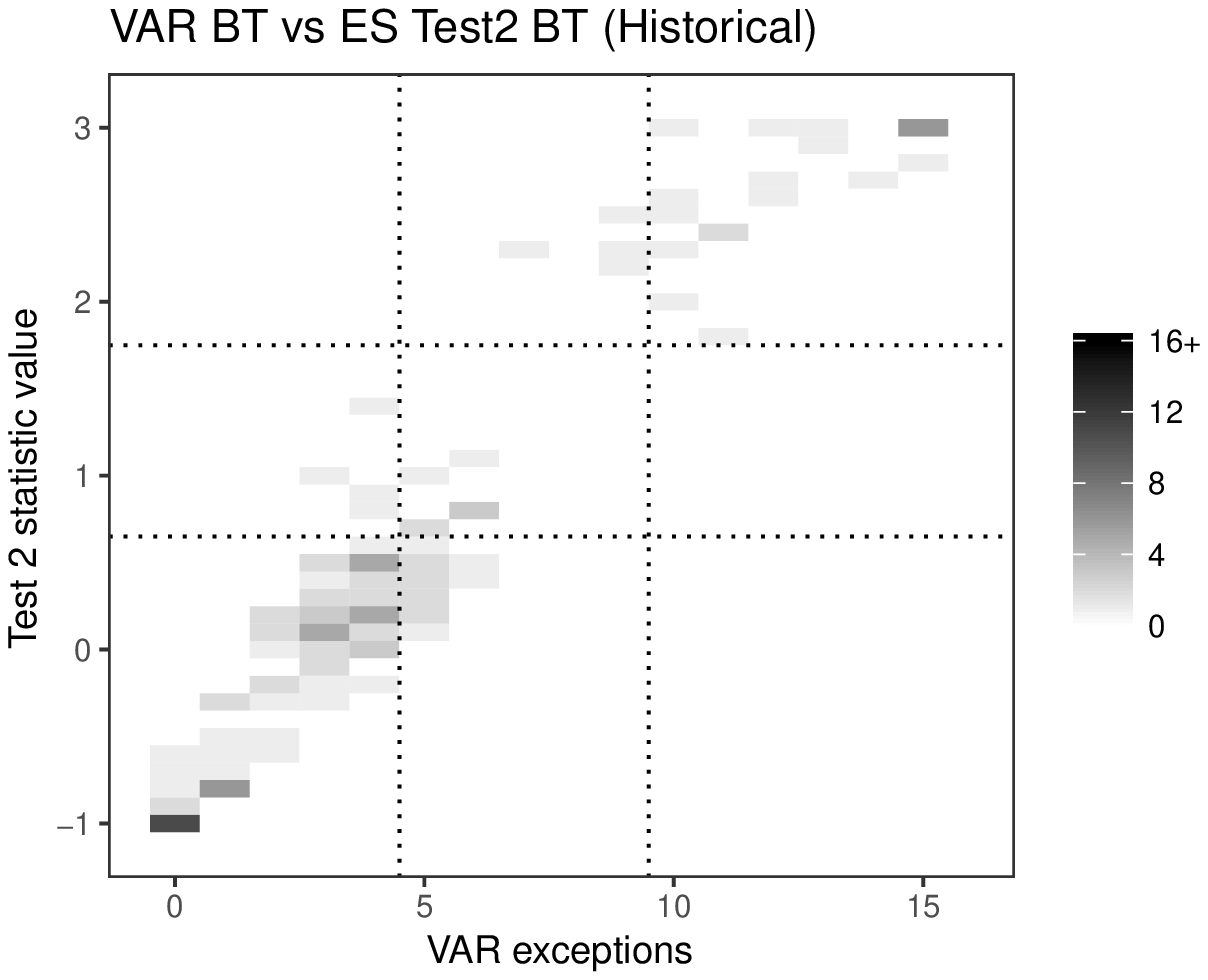}
\includegraphics[width=5cm]{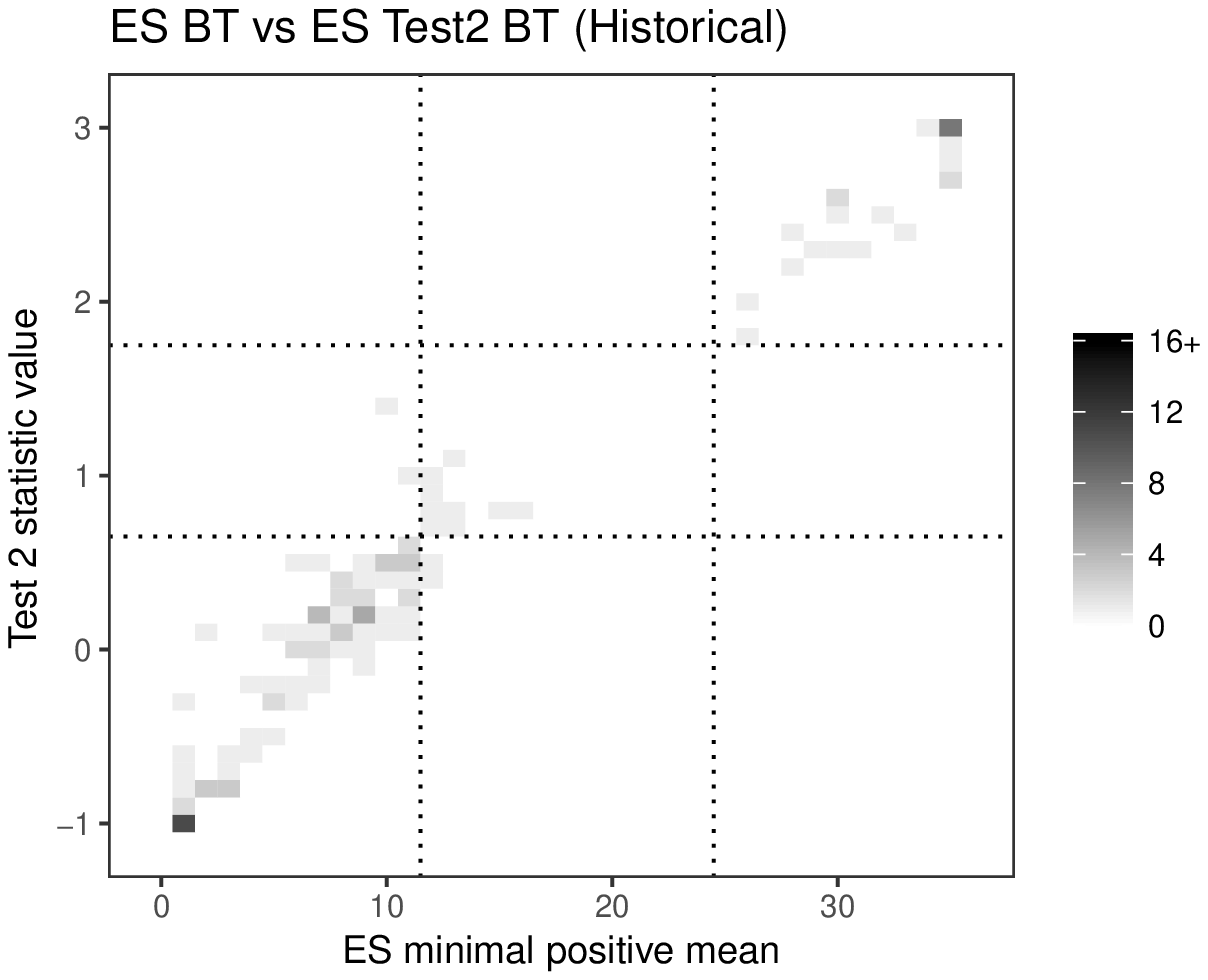}
\end{center}
\caption{Backtesting results for Market Data. For each sample $x$ we perform both VAR backtesting, ES backtesting and 'Test 2' from~\cite{Acerbi2014Risk}. We compare the results to see the consistency between our framework and 'Test 2'.}\label{F:test2.market}
\end{figure}

\begin{figure}[htp!]
\begin{center}
\includegraphics[width=5cm]{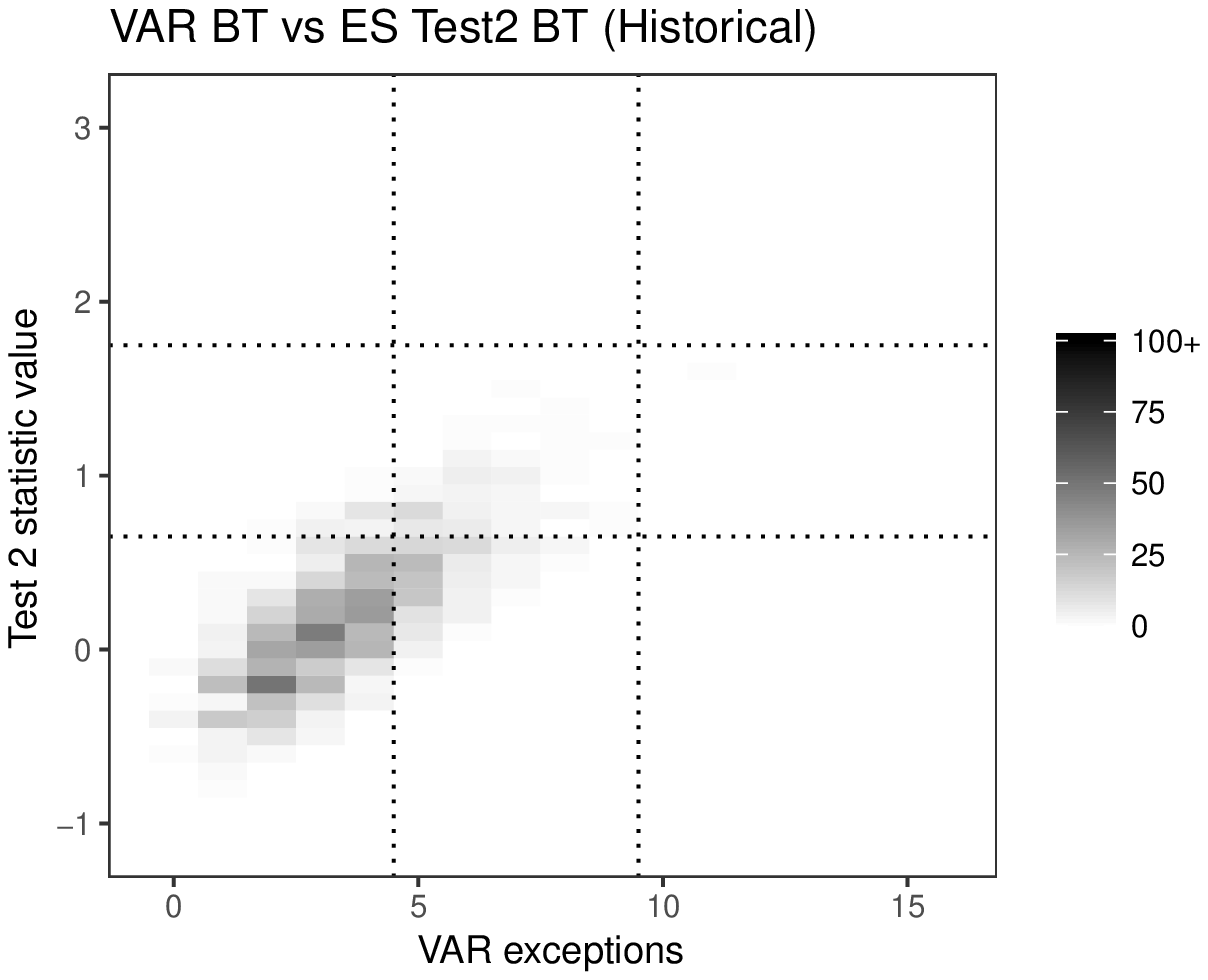}
\includegraphics[width=5cm]{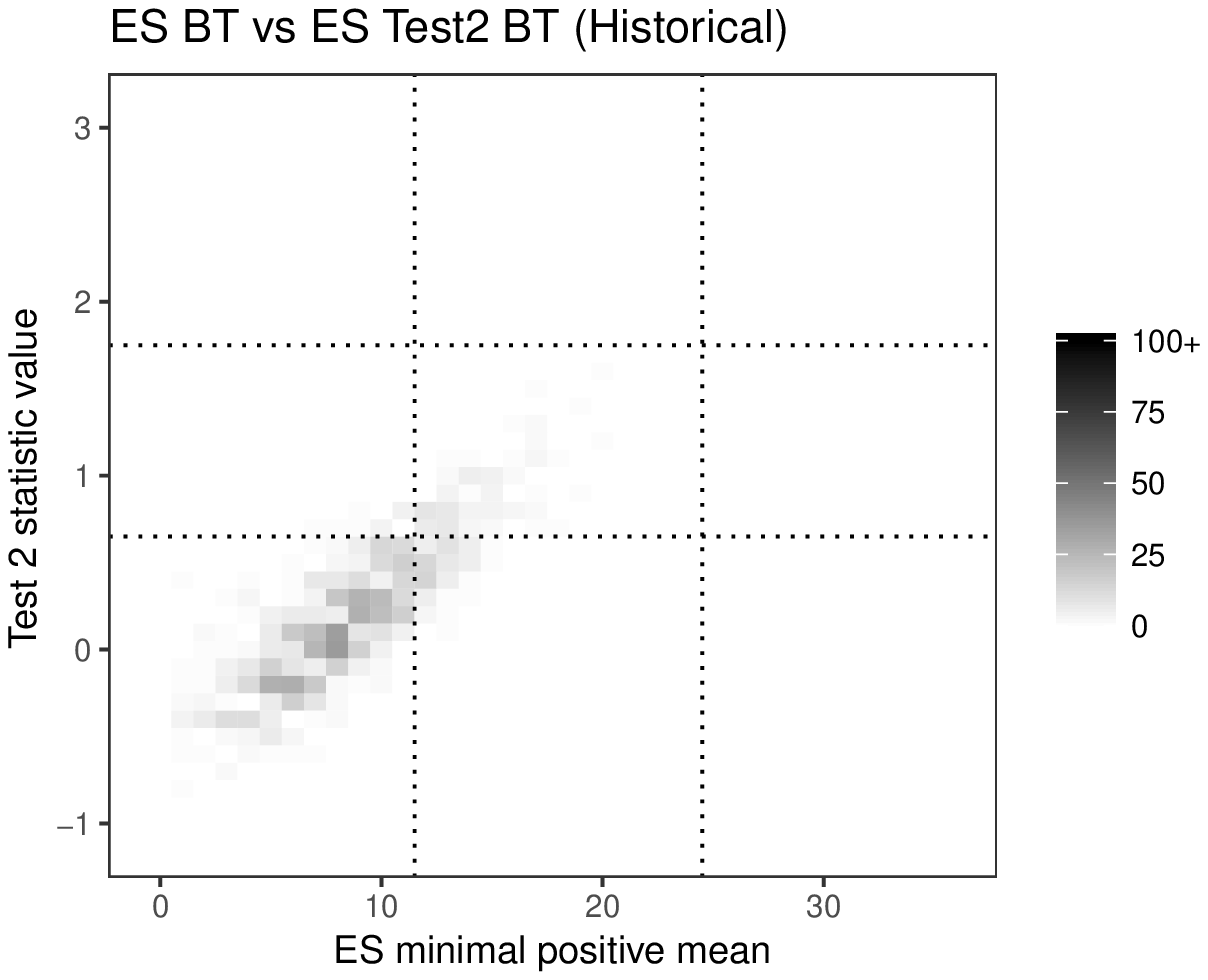}
\end{center}
\caption{Backtesting results for Normal Data. For each sample $x$ we perform both VAR backtesting, ES backtesting and 'Test 2' from~\cite{Acerbi2014Risk}. We compare the results to see the consistency between our framework and 'Test 2'.}\label{F:test2.normal}
\end{figure}

In Figure~\ref{F:test2.market} and Figure~\ref{F:test2.normal} we present the values of the test statistic $Z$ combined with the VAR exceptions as well as with ES worst-case positive sum scenarios for both market and normal data.

We see that {\it Test 2} is also consistent with the VAR framework (left plots) and consequently with our testing framework (right plots). Comparing Figure~\ref{F:market} with Figure~\ref{F:test2.market} (or Figure~\ref{F:normal} with Figure~\ref{F:test2.normal}) one can see that our framework gives a little bit more consistent results (the spread of the points is smaller) and that our framework is a little bit more conservative (e.g. there are more datasets with VAR yellow-zone classifications and 'test 2' green-zone classification) but the results are in fact very similar.

It should be noted that {\it Test 2} framework require IM methodology for both VAR and ES methodologies (in order to obtain test statistic $Z$) while our framework require only ES inputs (in order to get secured sample $y$).

\FloatBarrier

\section{Concluding remarks}\label{Sec:conclude}
In this short note we have introduced a new Expected Shortfall backtesting framework based on the duality between coherent risk measures and scale-invariant performance measures. The power of our backtesting framework proposal lies in its simplicity, elegance, and clear financial interpretation. Instead of calculating the number of breaches, as in the VAR framework, we focus on the biggest number of worst realisations for the secured position that add up to a negative total. We have shown that our framework is aligned with the current regulatory framework producing consistent classification schemes for correctly specified models. Also, contrary to many other proposed approaches, our framework is model independent, we do not need to consider joint $(\var,\text{ES})$ estimation and do not need to introduce the standard (reference, benchmark) estimation procedure for comparative backtesting in contrast to the elicitability-based backtests.

While our framework shares drawbacks that were also characteristic for the regulatory VAR model, we believe it could be used by the regulator and it's simplicity would make any potential manipulations easy to detect. Also, its transparency and clear financial interpretation makes the model very easy to describe and to implement in any programming environment. In {\bf R} software, using $y$ to denote the secured position sample vector, the value of $G_n$ could be computed in one line of code using the simple base functions: {mean(cumsum(sort(y))$<$0)}; to compute $T_n$ we can use code: {mean(y$<$0)}.

\appendix
\section{Proofs}\label{Sec:proofs}
\begin{proof}[Proof of Equality~\eqref{eq:t1.bis}]
Let us fix $n\in\bN$ and $y=(y_i)_{i=1}^{n}$. Using \eqref{eq:var.hist} and assuming that $y_{(n)}\geq 0$ we get
\begin{align*}
\inf \{\alpha\in (0,1):\, \hat\var^{\alpha}_{n}(y) \leq 0\} & =\inf \{\alpha\in (0,1):\, -y_{(\lfloor n\alpha\rfloor+1)} \leq 0\}\\
&=\frac{1}{n}\inf \{k\in (0,n):\, y_{(\lfloor k\rfloor+1)} \geq 0\}\\
&=\frac{1}{n}\inf \{k\in \{0,1,\ldots,n-1\}:\, y_{(k+1)} \geq 0\}\\
&=\frac{1}{n}\sum_{k=1}^{n}\1_{\{y_{(k)}<0\}}\\
&=\sum_{i=1}^{n} \frac{\1_{\{y_i<0\}}}{n}.
\end{align*}
On the other hand, if $y_{(n)}<0$ then using the convention $\inf\emptyset=1$ we conclude the proof.
\end{proof}
\begin{proof}[Proof of Equality~\eqref{eq:es.bis2}]
Let us fix $n\in\bN$ and $y=(y_i)_{i=1}^{n}$. Using \eqref{eq:es.hist} and assuming that $\sum_{i=1}^{n}y_{n}\geq 0$ we get
\begin{align*}
\inf \{\alpha\in (0,1):\, \hat{\textrm{ES}}^{\alpha}_{n}(y) \leq 0\} & =\inf \left\{\alpha\in (0,1):\, -\left(\frac{ \sum_{i=1}^{n}y_i\1_{\{y_i+\hat{\var}^{\alpha}_n(y)\leq 0\}}}{\sum_{i=1}^{n}\1_{\{y_i+\hat{\var}^{\alpha}_n(x)\leq 0\}}} \right) \leq 0\right\}\\
&=\inf \left\{\alpha\in (0,1):\, \sum_{i=1}^{n}y_i\1_{\{y_i+\hat{\var}^{\alpha}_n(y)\leq 0\}} \geq 0\right\}\\
&=\frac{1}{n}\inf \left\{k\in (0,n):\, \sum_{i=1}^{n}y_i\1_{\{y_i-y_{(\lfloor k\rfloor+1)} \leq 0\}} \geq 0\right\}\\
&=\frac{1}{n}\inf \left\{k\in \{0,1,\ldots,n-1\}:\, \sum_{i=1}^{n}y_{(i)}\1_{\{y_{(i)} -y_{(k+1)}\leq 0\}} \geq 0\right\}\\
&=\frac{1}{n}\inf \left\{k\in \{0,1,\ldots,n-1\}:\, \sum_{i=1}^{k+1}y_{(i)} \geq 0\right\}\\
&=\frac{1}{n}\sum_{k=1}^{n}\1_{\{y_{(1)}+\ldots+y_{(k)}<0\}}.
\end{align*}
On the other hand, if $\sum_{i=1}^{n}y_{n}>0$ then using the convention $\inf\emptyset=1$ we conclude the proof.
\end{proof}

 \end{document}